\documentclass[longauth]{aa}  

\usepackage{graphicx}
\usepackage{txfonts}
\usepackage{subfig}
\begin{document}

   \title{Stellar metallicity from optical and UV spectral indices: Test case for WEAVE-StePS}

   \author{F. R. Ditrani
          \inst{1,2}, M. Longhetti\inst{1}, F. La Barbera\inst{3}, A. Iovino\inst{1}, L. Costantin\inst{4}, S. Zibetti\inst{5}, A. Gallazzi\inst{5}, M. Fossati\inst{1,2}, J. Angthopo\inst{1}, Y. Ascasibar\inst{6}, B. Poggianti\inst{7}, P. Sánchez-Blázquez\inst{6,8}, M. Balcells\inst{9,10,11}, M. Bianconi\inst{12}, M. Bolzonella\inst{13}, L. P. Cassarà\inst{14}, O. Cucciati\inst{13}, G. Dalton\inst{15,16}, A. Ferré-Mateu\inst{10,11}, R. García-Benito\inst{17}, B. Granett\inst{1}, M. Gullieuszik\inst{7}, A. Ikhsanova\inst{18}, S. Jin\inst{15,19,20} , J. H. Knapen\inst{10,11}, S. McGee\inst{12}, A. Mercurio\inst{3,21}, L. Morelli\inst{22,1}, A. Moretti\inst{7}, D. Murphy\inst{23}, A. Pizzella\inst{18,7}, L. Pozzetti\inst{13}, C. Spiniello\inst{15,3}, C. Tortora\inst{3}, S. C. Trager\inst{19}, A. Vazdekis\inst{10,11}, D. Vergani\inst{13}, and B. Vulcani\inst{7}
          }

   \institute{INAF-Osservatorio Astronomico di Brera, via Brera 28, I-20121 Milano, Italy\\%1
              \email{fabio.ditrani@inaf.it}
    \and{Università degli studi di Milano-Bicocca, Piazza della scienza, 20125 Milano, Italy}%2
    \and {INAF - Osservatorio Astronomico di Capodimonte, Via Moiariello 16, I-80131 Napoli, Italy}%3
    \and {Centro de Astrobiología (CSIC-INTA), Ctra de Ajalvir km 4, Torre- jón de Ardoz, E-28850, Madrid, Spain}%4
    \and {INAF - Osservatorio Astrofisico di Arcetri, Largo Enrico Fermi 5, I-50125 Firenze, Italy}%5
    \and {Departamento de Física Teórica, Universidad Autónoma de Madrid, E-28049 Madrid, Spain}%6
    \and{INAF – Osservatorio Astronomico di Padova, Vicolo dell’Osservatorio 5, I-35122 Padova, Italy}%7
    \and{Instituto de Física de Partículas y del Cosmos (IPARCOS), Universidad Complutense de Madrid, E-28040 Madrid, Spain}%8
    \and {Isaac Newton Group of Telescopes, ING, 38700 La Palma (S.C. Tenerife), Spain}%9
    \and{Instituto de Astrofísica de Canarias, IAC, Vía Láctea s/n, E-38205, La Laguna (S.C. Tenerife), Spain}%10
    \and{Departamento de Astrofísica, Universidad de La Laguna, E-38206, La Laguna (S.C. Tenerife), Spain}%11
    \and {School of Physics and Astronomy, University of Birmingham, Birmingham, B15 2TT, UK}%12
    \and {INAF – Osservatorio di Astrofisica e Scienza dello Spazio, Via P. Gobetti 93/3, I-40129 Bologna, Italy}%13
    \and{INAF - IASF Milano, via Bassini 15, I-20133 Milano, Italy} %14
    \and {Dept. Physics, University of Oxford, Keble Road, Oxford OX1 3RH, U.K.}%15
    \and {RAL, Space, Science and Technology Facilities Council, Harwell, Didcot OX11 0QX, U.K.}%16
    \and {Instituto de Astrofísica de Andalucía (CSIC), P.O. Box 3004, E-18080, Granada, Spain}%17
    \and {Dipartimento di Fisica e Astronomia “G. Galilei”, Università di Padova, vicolo dell’Osservatorio 3, I-35122, Padova, Italy}%18
    \and{Kapteyn Astronomical Institute, Rijksuniversiteit Groningen, Landleven 12, 9747 AD Groningen, the Netherlands}%19
    \and{SRON - Netherlands Institute for Space Research, Landleven 12, 9747 AD Groningen, the Netherlands}%20
    \and{Dipartimento di Fisica “E.R. Caianiello”, Università degli studi di Salerno, Via Giovanni Paolo II 132, I-84084 Fisciano (SA)}%21
    \and{Instituto de Astronomía y Ciencias Planetarias de Atacama (INCT), Universidad de Atacama, Copayapu 485, Copiapó, Chile}%22
    \and {Institute of Astronomy, University of Cambridge, Madingley Road, Cambridge CB3 0HA, U.K.}}%23

   \date{Received; accepted}

% \abstract{}{}{}{}{} 
% 5 {} token are mandatory
 
  \abstract
   {The upcoming generation of optical spectrographs on four meter-class telescopes, with their huge multiplexing capabilities, excellent spectral resolution, and unprecedented wavelength coverage, will provide high-quality spectra for thousands of galaxies. These data will allow us to  examine of the stellar population properties at intermediate redshift, an epoch that remains unexplored by large and deep surveys.}
   {We assess our capability to retrieve the mean stellar metallicity in galaxies at different redshifts and signal-to-noise ratios (S/N), while simultaneously exploiting the ultraviolet (UV) and optical rest-frame wavelength coverage.}
   {The work is based on a comprehensive library of spectral templates of stellar populations, covering a wide range of age and metallicity values and built assuming various star formation histories (SFHs),  to cover an observable parameter space with diverse chemical enrichment histories and dust attenuation. We took into account possible observational errors, simulating realistic observations of a large sample of galaxies carried out with WEAVE at the William Herschel Telescope at different redshifts and S/N values.  We measured all the available and reliable indices on the simulated spectra and on the comparison library. We then adopted a Bayesian approach to compare the two sets of measurements in order to obtain the probability distribution of stellar metallicity with an accurate estimate of the uncertainties.}
   {The analysis of the spectral indices has shown how some mid-UV indices, such as BL$3580$ and Fe$3619$, can provide reliable constraints on stellar metallicity, along with optical indicators.
   The analysis of the mock observations has shown that even at S/N = 10, the metallicity can be derived within $0.3$ dex, in particular, for stellar populations older than $2$ Gyr. The S/N value plays a crucial role in the uncertainty of the estimated metallicity and so, the differences between S/N = 10 and S/N = 30 are quite large, with uncertainties of $\sim 0.15$ dex in the latter case. On the contrary, moving from S/N = 30 to S/N = 50, the improvement on the uncertainty of the metallicity measurements is almost negligible. Our results are in good agreement with other theoretical and observational works in the literature and show how the UV indicators, coupled with classic optical ones, can be advantageous in constraining metallicities.}
   {We demonstrate that a good accuracy can be reached on the spectroscopic measurements of the stellar metallicity of galaxies at intermediate redshift, even at low S/N, when a large number of indices can be employed, including some UV indices. This is very promising for the upcoming surveys carried out with new, highly multiplexed, large-field spectrographs, such as StePS at the WEAVE and 4MOST, which will provide spectra of thousands of galaxies covering large spectral ranges (between $3600$ and $9000 \ \AA$ in the observed frame) at relatively high S/N ($> 10 \ \AA^{-1}$).}

   \keywords{galaxies: evolution -- galaxies: formation -- galaxies: stellar content
               }
   \titlerunning{Stellar metallicity from optical and UV spectral indices}
\authorrunning{Ditrani et al.}
   \maketitle

%
%-------------------------------------------------------------------

\section{Introduction}

Tracing stellar ages, chemical abundances, and masses is a very powerful way to probe the evolution of galaxies, and to explore the physical mechanisms of galaxy assembly \citep{conroy2013modeling}. The metal content, in particular, holds the imprint of the baryonic cycle that regulates the star formation in galaxies by balancing the inflow of pristine gas, the outflow of metal-loaded gas blown out by stellar and active galactic nucleus (AGN) winds, and the re-accretion of this metal-enriched gas \citep{pagel1975metal,peeples2011constraints,peng2014dependence,hunt2020scaling,tortora2022scaling}.
While the metallicity obtained from nebular emission lines refers to the gas component, the one determined from stellar continuum and absorption lines refers to the stellar component.
One of the most direct ways to obtain information about the metal content of the stellar populations in galaxies is to compare the stellar spectra of galaxies with synthetic templates based on evolutionary stellar population synthesis (SPS) models \citep[e.g.][]{bruzual1993spectral,bruzual2003stellar,maraston1998evolutionary,vazdekis1999evolutionary,maraston2011stellar,vazdekis2016uv}. 
However, obtaining such an estimate is a challenging task, due to the degeneracy between the different stellar populations parameters, namely, the age, metallicity, dust content, and IMF slope of the galaxy.
%One of the main challenges of retrieving the stellar population parameters is the degeneracy between the age, metallicity and dust content of the galaxy. 
In fact, changes in any of these parameters can produce similar effects on the resulting spectrum. Comparisons between models and observations can be done for selected spectral absorption features that are most sensitive to the stellar populations parameters \citep[e.g.][]{worthey1994old,worthey1997hgamma,vazdekis1997new,thomas2005epochs,gallazzi2005ages,spiniello2014stellar,la2017imf,conroy2018metal,maraston2020stellar}. Indeed, the advantage of using line-strength indices is the possibility to select specific spectral features that are mainly sensitive to the age or to the metallicity of the stellar populations. Moreover, using only narrow regions of the whole spectrum drastically reduces the effect of dust attenuation, whose main effect is on the spectrum overall shape.
Spectral indices are thus one of the most effective tools to derive stellar metallicity in galaxies. However, high-quality and at least moderate-resolution ($R \ge 2000$) spectra are needed to measure the absorption indices that are sensitive to the age and metallicity of the stellar populations, since many metallic absorption lines are shallow and narrow and thus difficult to measure.

In the last few decades, studies of Local Universe galaxies have greatly enriched our understanding of galaxy evolution. A large number of high-quality spectra from the Sloan Digital Sky Survey \citep[SDSS,][]{york2000sloan} have provided a robust anchor both to theoretical and empirical approaches at $z \sim 0$. For example, \cite{gallazzi2005ages} found that low-mass galaxies are on average younger and more metal-poor than higher-mass galaxies.
However, evolutionary scenarios are degenerate, since different formation and evolution tracks can result in the same galaxy population at $z \sim 0$. Therefore, spectroscopic studies at higher redshift are needed to better understand the evolution of galaxies during cosmic time.
Studies on stellar populations in galaxies at higher redshift are observationally challenging as they require spectroscopy with a high signal-to-noise ratio (S/N $> 10 \ \AA^{-1}$) and moderate resolution to trace the key absorption indices that are sensitive to age and metallicity. Moreover, the absorption line indices historically adopted to derive the stellar metallicity in local galaxies are mainly located in the optical rest-frame region, which move to redder wavelengths at higher
redshift, that are strongly affected by sky emission lines and telluric absorptions. Indeed, medium-and high-redshift studies at the quality level required to derive stellar parameters of galaxies are limited to few spectroscopic works, mainly based on cluster galaxies at $z < 1$ \citep[e.g.][]{jorgensen2005rx,sanchez2009evolution,jorgensen2013stellar,ferre2014tale}, with some exceptions in the field \citep[e.g.][]{ferreras2009early,gallazzi2014charting}.
The recently completed LEGA-C public survey \citep{van2016vlt,straatman2018large,van2021large} is a first step in this direction, since it has gathered S/N $\sim 20$ spectra for $\sim 3000$ galaxies at $0.6 < z < 1.0$, suitable for obtaining statistically robust characterisations of the stellar population parameters  in this redshift range for the first
time \citep[e.g.][]{wu2018fast,chauke2018star,d2020inverse}. Surprisingly, the intermediate redshift window $0.3 < z < 0.6$, crucial to link the high redshift observations with the Local Universe galaxies, remains unexplored due to the lack of large surveys with high-quality spectra. 

The upcoming generation of spectrographs on four-meter-class telescopes, with their huge multiplexing capabilities, wide wavelength coverage, and moderate spectral resolution, can offer an interesting opportunity to fill in this redshift gap. In particular, the new spectrographs will provide spectra with similar quality of those from SDSS in the Local Universe. Two complementary ambitious surveys of $0.3 < z < 0.7$ massive galaxies will start soon: the WEAVE-StePS project on the $4.2$m William Herschel Telescope in the Canary Islands \citep[WHT,][]{dalton2012weave,jin2023wide,iovino2023} and the accepted $4$MOST-StePS at $4.1$m ESO-VISTA in Paranal (Messenger, in prep.). %In this paper we focus on the new wide-field WEAVE spectrograph, located at the $4$mt class William Herschel Telescope in La Palma \citep[][Jin et al. 2022, \textit{in press}]{dalton2012weave}, and on the Stellar Population at intermediate redshift Survey (StePS, Iovino et al., in prep.). The survey which will start at the beginning of $2023$ and will last $5$ years. 
Thanks to their new-generation spectrographs, both surveys will provide  spectra (with S/N $> 10\AA^{-1}$) for  hundreds of galaxies in one shot at $R = 5000$, in the observed spectral range from $3660$ to $9590\AA$.
The two surveys will provide, for the first time, thousands of high-quality spectra of galaxies, selected to have $I_{AB} \le 20.5$ mag and photometric (as well as spectroscopic, when available) redshift  of $0.3 <z < 0.7$, in order to enable a continuous reconstruction of the evolutionary path of galaxies from $z \sim 1$ to the Local Universe. The magnitude cut implies a stellar mass limit of galaxies of $M\sim 10^{10.4} M_\odot$ at $z = 0.3$, $M\sim 10^{11} M_\odot$ at $z = 0.55$ and $M\sim 10^{11.3} M_\odot$ at $z = 0.7$ \citep[assuming a pure passive evolution, see][]{iovino2023}, to sample the massive tail of the galaxy population. In particular, the WEAVE-StePS survey will observe a large sample ($\sim 25000$ galaxies) at S/N $\sim 10$, suitable for measuring stellar population parameters. The 4MOST-StePS will focus instead on a smaller yet representative sample of galaxies, similar in size to LEGA-C, trading the sample size for a much higher S/N ($\sim 30$).

The goal of our paper is to exploit the information provided by key UV and optical absorption-line indices coupled with a Bayesian approach in order to infer the mass-weighted stellar metallicity of galaxies. To achieve this purpose, we created realistic simulations of WEAVE-StePS spectra at different redshifts and S/N. Thanks to the wide wavelength range provided by WEAVE, we are able to study a large number of spectral absorption-line indices both in the UV and optical regions. 

The structure of the paper is as follows. In Sect.~\ref{sec:model}, we describe the stellar population models used in this work. In Sect.~\ref{sec:index}, we explore the possibility to infer the metallicity using information from pairs of individual optical and UV indices. In Sect.~\ref{sec:stepsobs}, we describe our method to simulate realistic WEAVE-StePS-like observations, which closely mimic spectra that will be observed by WEAVE. In Sect.~\ref{sec:analysis}, we describe the analysis we carried out to obtain a robust measurement of metallicity with a Bayesian approach. In Sect.~\ref{sec:discussion}, we present our ability to infer the metallicity and we compare it to the literature. In Sect.~\ref{sec:bluer}, we present a more generic test to explore the capability of retrieving the stellar metallicity using different set of indices.
In Sect.~\ref{sec:conclusion}, we summarise our results and we present our conclusions and future applications.
Throughout, we adopt a standard $\Lambda$CDM cosmology with $\Omega_M = 0.286$, $\Omega_\Lambda = 0.714,$ and $H_0 = 69.6$ km s$^{-1}$ Mpc$^{-1}$ \citep{wright2006cosmology,bennett20141}. Magnitudes are given in the AB system \citep{oke1974absolute}. 
%--------------------------------------------------------------------
\section{Stellar population models}
\label{sec:model}

Following \cite{costantin2019few}, we used a comprehensive library of spectral templates of stellar populations \citep{zibetti2017resolving}. The library is based on a revised version of the BC03 models by \cite{bruzual2003stellar}.\footnote{
The models used in \cite{costantin2019few} were a CB$16$ version, while the models adopted in this work are the official $2019$ public release (\textit{http://www.bruzual.org/CB19/}), referred as C\&B in \citealt{sanchez2022sdss} (private communication by G. Bruzual).}

%The adopted version is a major revision of the BC03 stellar population sythesis models, introduced in \cite{plat2019constraints}(-Appendix A, referred as C\&B models). 
The new version follows the PARSEC evolutionary tracks \citep[][]{marigo2013evolution,chen2015parsec} for $16$  metallicity values, assuming a solar abundance of $Z_\odot = 0.017$. The new tracks include evolution of stars from the main sequence stage to the Wolf-Rayet phase for the most massive stars or to the thermally pulsing asymptotic giant branch (TP-AGB) for stars with mass lower than $6 M_\odot$. Details on the description of the Wolf-Rayet phase adopted in the models can be found in Appendix A of \cite{plat2019constraints}. For a more detailed description of the new adopted ingredients of the models (cited as C\&B), we refer to Appendix A of \cite{sanchez2022sdss}.

We assumed a Chabrier initial mass function \citep[IMF,][]{chabrier2003galactic} with $M_{UP} = 100 M_\odot$ and the MILES stellar library \citep[$3540.5 < \lambda < 7350.2 \AA$;][$2.5\AA$ FWHM resolution]{sanchez2006medium,falcon2011updated}, extended in the UV ($911 < \lambda < 3540.5 \AA$) with theoretical high-resolution models \citep[][$1\AA$ FWHM, see also Table 12 in \citealt{sanchez2022sdss}]{martins2005high} . 
Each simple stellar population (SSP) model provides $220$ spectra computed at different time steps ranging from $0.01$ Myr to $14$ Gyr, with a metallicity ranging from $-1.7$ dex to $0.4$ dex \citep[more details in][]{plat2019constraints,sanchez2022sdss}.
The templates in our library have been built assuming different SFHs, chemical enrichment histories, and dust attenuation values \citep[following the two components attenuation model of][]{charlot2000simple}, covering extensively the space of observables. The SFHs are composed of a secular component, described by a \cite{sandage1986star} law:
\begin{equation}
    \textrm{SFR}_\tau(t) \propto \frac{t}{\tau}\exp\left({-\frac{t^2}{2\tau^2}}\right)
,\end{equation}
with a superposition of random bursts (in two-thirds of the templates). The total mass formed in these bursts ranges between $1/1000$ and two times the total stellar mass formed in the secular component.
The resulting templates library roughly uniformly covers the mean mass-weighted log age-Z plane, between $-1.70 < \mathrm{[Z/H]} < 0.4$ in the metallicity range, where solar metallicity is [Z/H]$_{\text{sol}} = 0$, and $8.5< \log(\text{age yr}^{-1}) < 10.2$ in the age range. 
All the galaxy types are well represented, from the star forming galaxies to the quiescent ones, with the only exception of the starburst galaxies at their very early stage (i.e. characterised by high star formation rate and very small mass formed).
More details about these templates can be found in \cite{zibetti2017resolving} and \cite{costantin2019few}. 
We used a representative subset of the above library to generate mock WEAVE-StePS-like spectroscopic observations, while the remaining ones have been used as a comparison library to recover the input physical parameters using the Bayesian statistical tools described in Sect.~\ref{sec:analysis} \citep[see also][]{gallazzi2005ages}.

\section{Optical and UV spectral indices: A direct comparison}
\label{sec:index}
%----------------------------------------------------------------- 
   \begin{figure*}
   \centering
   \includegraphics[width=0.95\textwidth]{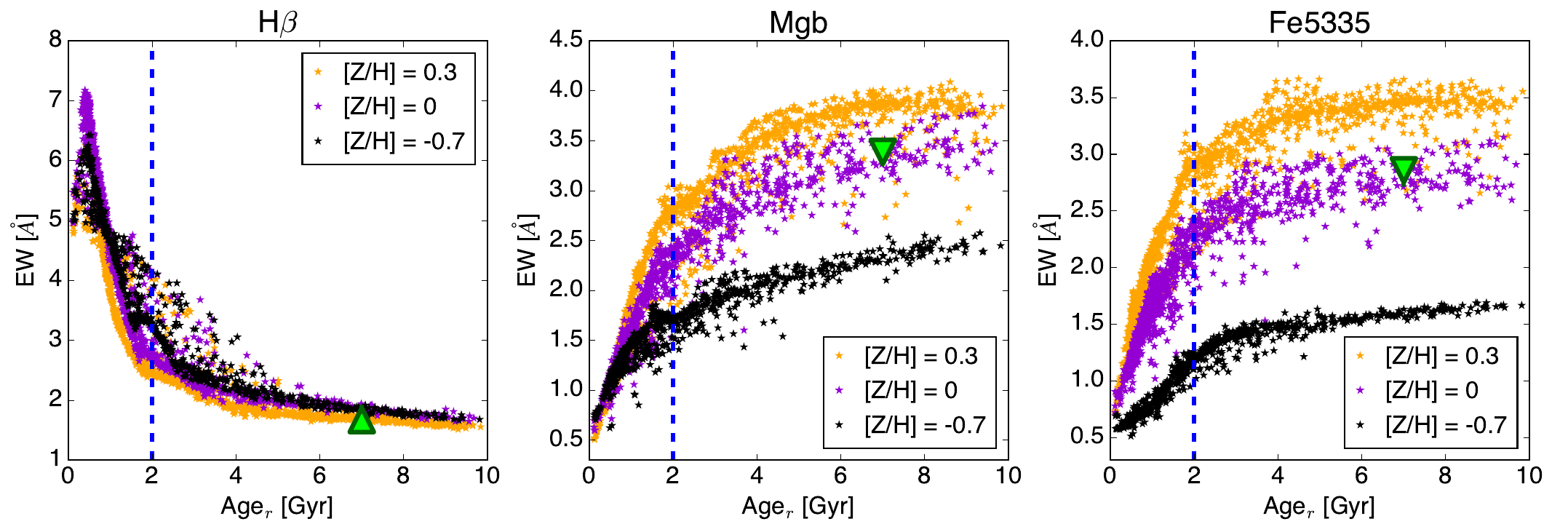}
      \caption{Distribution of H$\beta$ (left panel), Mgb (middle panel) and Fe$5335$ (right panel) values as a function of the $r$-band light-weighted age in three different bins of mass-weighted [Z/H]: $0.3$ (orange symbols), $0$ (purple symbols), and $-0.7$ (black symbols), respectively, as measured in our template library (excluding templates with secondary bursts). The green arrow represents the effect on the indices of a $0.1\%$ fraction of a young ($70$ Myr) stellar population superimposed on a population with $7$ Gyr, both with a solar metallicity. 
              }
         \label{fig:mgbbl}
   \end{figure*}
%------------------

%----------------------------------------------------------------- 
   \begin{figure*}
   \centering
   \includegraphics[width=0.95\textwidth]{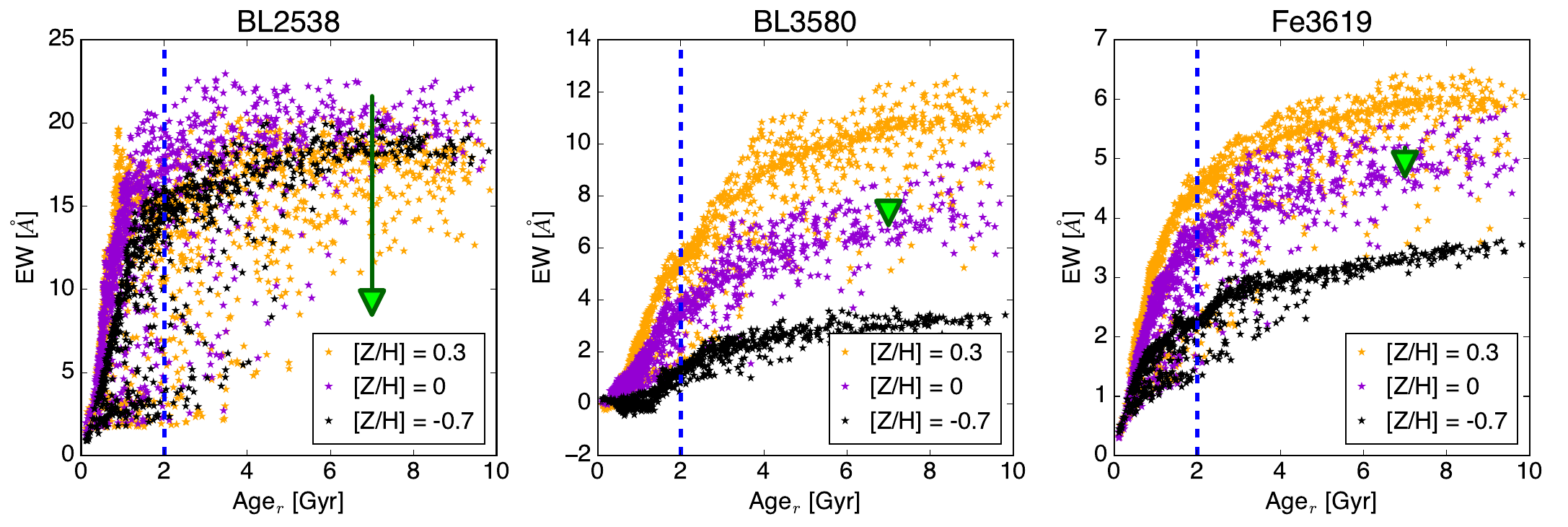}
      \caption{Same as Figure~\ref{fig:mgbbl}, but for BL$2538$ (left panel), BL$3580$ (middle panel), and Fe$3619$ (right panel), respectively.
              }
         \label{fig:blfe}
   \end{figure*}
%------------------

%-------------------------------------- Two column figure (place early!)
   \begin{figure*}
   \centering
   \subfloat[][\emph{}]
        {\includegraphics[width=0.45\textwidth]{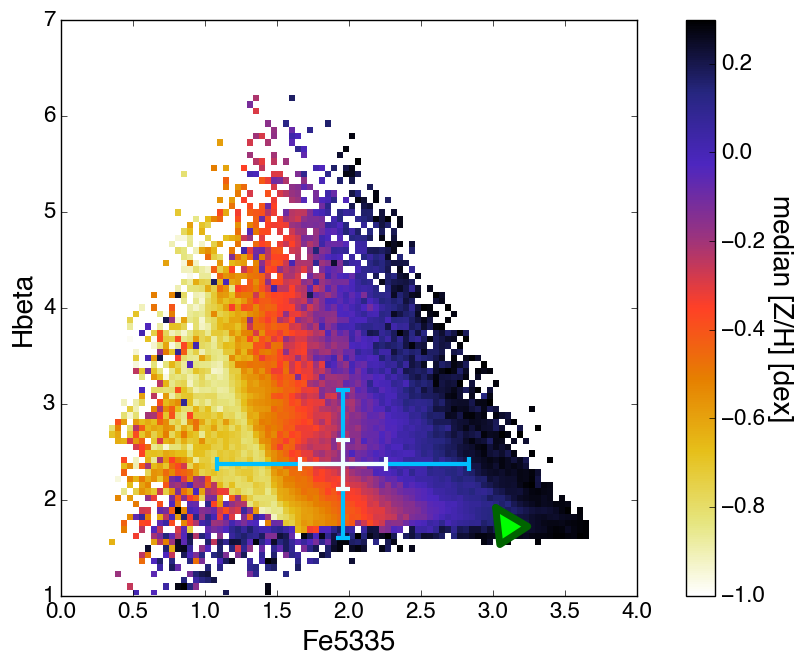}} \quad
        \subfloat[][\emph{}]
        {\includegraphics[width=0.45\textwidth]{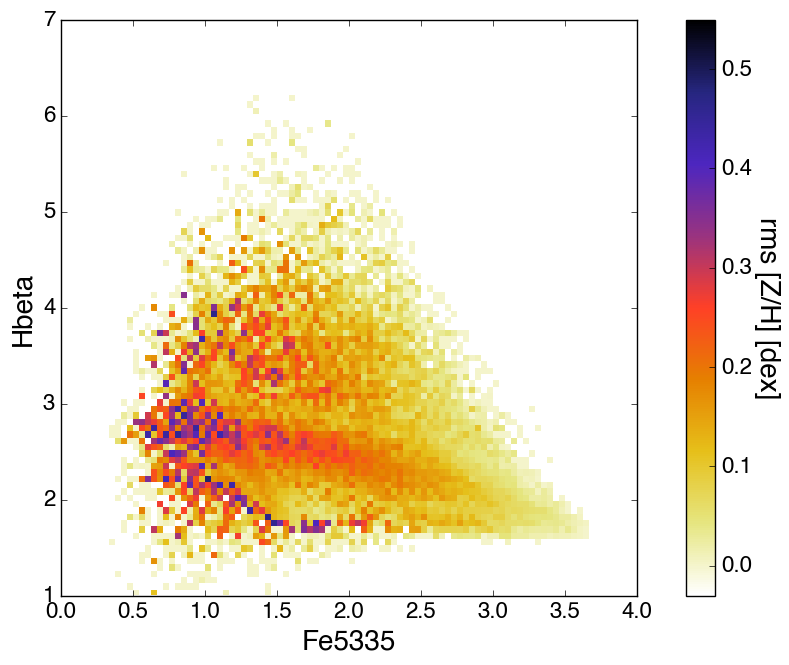}} \quad
 
   \caption{(Fe$5335$,H$\beta$) diagnostic plane. Left panel: (Fe$5335$,H$\beta$) diagnostics colour-coded according to metallicity for the templates library, considering galaxies with $r$-band light-weighted age $> 2$ Gyr. Right panel: (Fe$5335$,H$\beta$) diagnostics colour-coded according to the rms in metallicity. The light blue and white error bars represent the typical median $1\sigma$ error in measuring Fe$5335$ and H$\beta$ at S/N$_{I,\textrm{obs}} = 10$ and S/N$_{I,\textrm{obs}} = 30$, respectively. The green arrow represents the effect on the indices of a $0.1\%$ fraction of a stellar population with $70$ Myr superimposed on a population with $7$ Gyr, each with solar metallicity.} 
              \label{fig:mgbhbeta}
    \end{figure*}

   \begin{figure*}
   \centering
   \subfloat[][\emph{}]
        {\includegraphics[width=0.45\textwidth]{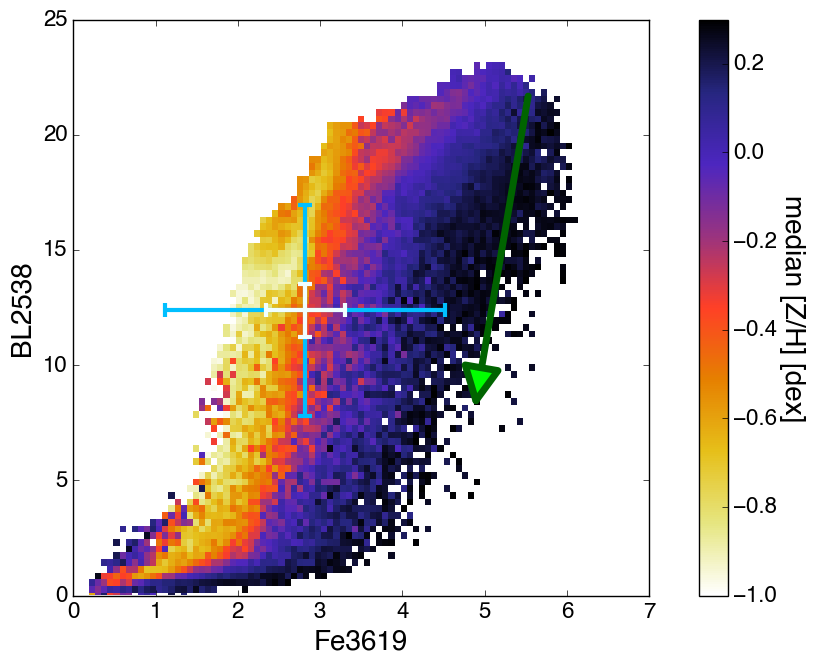}} \quad
        \subfloat[][\emph{}]
        {\includegraphics[width=0.45\textwidth]{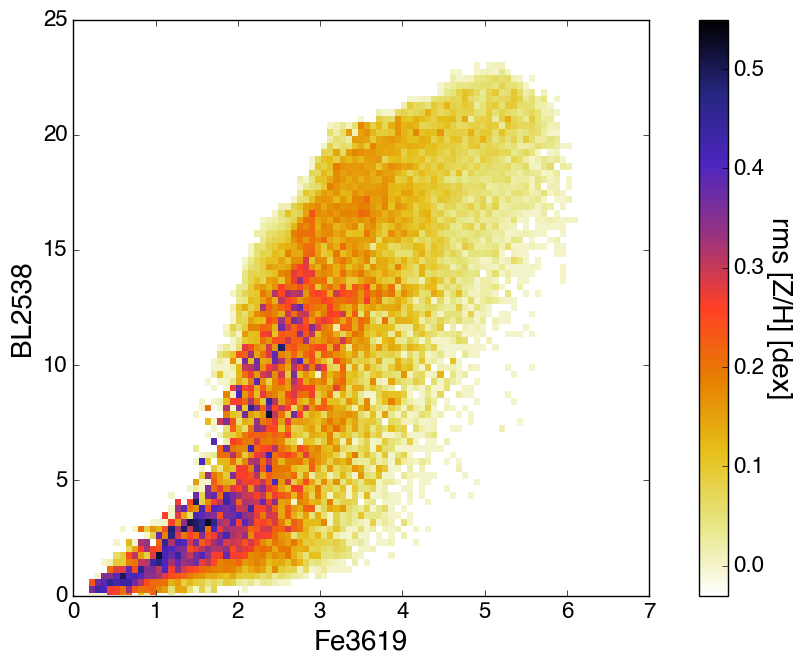}} \quad
 
   \caption{Same as Figure~\ref{fig:mgbhbeta}, but for (Fe$3619$,BL$2538$).}
              \label{fig:2bl}
    \end{figure*}

%--------------------------------------------------------------------

In the literature, stellar metallicity is usually determined by measuring the equivalent width of optical absorption line-strengths with respect to pseudo-continua, as defined in \cite{worthey1994old}. In particular, specific optical spectral indices such as Mgb and Fe$5335$, combined with age-sensitive indices (e.g. H$\beta$), help to break the age-metallicity degeneracy \citep[e.g.][]{boardman2017integral,sarzi2018fornax3d}. 
Figure~\ref{fig:mgbbl} shows the distribution of H$\beta$, Mgb, and Fe$5335$ values as a function of the $r$-band light-weighted age for three narrow bins of mass-weighted metallicities, [Z/H], measured in our templates library, but excluding cases with secondary bursts. 
As expected, H$\beta$ values show a strong dependence on the $r$-band light-weighted age of the galaxy stellar populations, while Mgb and Fe$5335$ reveal a clear metallicity dependence, in addition to the age dependency. In particular the two metallicity-sensitive indices display similar behaviour, with a strong stellar metallicity dependence for ages older than $2$ Gyr, whereas the age dependency nearly disappears. Typically, metallicity indices start to be mainly sensitive to the presence of hot stars for ages younger than $2$ Gyr, implying a dependence on stellar age in that regime. This is particularly true for Mgb, whose absorption is prominent in the atmosphere of giant stars.

In recent decades, UV spectral indices, also sensitive to the stellar age and metallicity, have been used to infer stellar population properties in galaxies \citep[e.g.][]{fanelli1992spectral,daddi2005passively,maraston2009absorption,vazdekis2016uv,lonoce2020stellar,salvador2021young}. %Upcoming facilities such as WEAVE and 4MOST will soon offer the possibility to explore the ultraviolet region from $2500\AA$ to $3800\AA$ and thus exploit UV indices.
Figure~\ref{fig:blfe} shows, as an example, the dependence of three UV indices, namely, the age-sensitive index BL$2538$ and the metallicity-sensitive indices BL$3580$ and Fe$3619$, on $r$-band light-weighted age. As for Mgb and Fe$5335$, the indices BL$3580$ and Fe$3619$ show a prominent dependence on [Z/H] for ages older than $2$ Gyr, while the BL$2538$ index depends on both age and metallicity.
For a more complete understanding of the behaviour of the UV indices (and continuum), particularly below $3000\AA$, as function of age and metallicity, it should be noted that they can be strongly affected by the presence of even a tiny fraction (i.e. $< 0.1\%$ of the overall mass) of very young (i.e. age $< 100$ Myr) and hot stars, with drastic changes seen in the continuum and spectral features \citep{cassara2013role,vazdekis2016uv,salvador2020sub,salvador2021young,corcho2021galaxies,salvador2022lessons}. 
In our simulations cases of quiescent galaxies with a tiny fraction of a very young populations are rare; however, they seem quite common, at least in the Local Universe, therefore we explored the behaviour of UV and optical indices considering a combination of two SSPs model.
In Figures~\ref{fig:mgbbl} and~\ref{fig:blfe}, the green arrows indicate the effect of adding $0.1\%$ of mass fraction of a $70$ Myr SSP to an old SSP of $7$ Gyr, both at solar metallicity. As we can see from Figs.~\ref{fig:mgbbl} and~\ref{fig:blfe}, even a tiny fraction of a very young population has an important effect in the spectral region below $3000\AA$. Strong effect on the UV wavelength region of rather old stellar
populations can be produced also by old hot stars such as the
post-asymptotic giant branch (PAGB) (e.g.  \citealt{le2016modelling}), whose
treatment in stellar population models is rather uncertain \citep[e.g.  due to the
effect of mass-loss on advanced phases of stellar evolution; see][]{Ma2006pagb,conroy2013modeling}, or those on the blue horizontal branch (not included in our templates). On the contrary, the effect is drastically reduced for indices in the wavelength region around $3500\AA$, and becomes even more negligible when moving to optical indices. 
%that, therefore, have a good, effective, sensitivity to stellar metallicity.

%Figure~\ref{fig:blfe} shows the distribution of three indices values in the considered UV region, BL$2538$, BL$3580$ and Fe$3619$, as a function of the $r$-band light-weighted age at three different [Z/H] for the model library, measured in our models library but excluding the more complex SFH which include secondary bursts. 

%Figure~\ref{fig:blfe} shows the age sensitivity of the BL$2538$ index and the metallicity dependence of BL$3580$ and Fe$3619$. Again, the two metallicity sensitive indices show their sensitivity from ages older than $2$ Gyr, as the two optical indices Mgb and Fe$5335$.

Classically, index-index diagnostics, which combine an age and a metallicity index indicator, are used to infer the metallicity (and the age) of stellar populations to reduce the age-metallicity degeneracy \citep[e.g.][]{trager2000stellar,longhetti2000star}. The (Fe$5335$, H$\beta$) or (Mgb, H$\beta$) diagnostic diagrams have been often used to constrain the metallicity from low-to high-redshift galaxies \citep[see][and references therein]{cervantes2009optimized}. 
Figure~\ref{fig:mgbhbeta} shows the (Fe$5335$, H$\beta$) diagnostic plane colour-coded according to the median mass-weighted metallicity (left panel) and to the intrinsic scatter of the metallicity (right panel) for the library previously described, including secondary bursts but considering galaxies with $r$-band light-weighted age $> 2$ Gyr. 
The different metallicity values occupy different regions of the index-index plane, with some mixing of values restricted only in the lower left corner (i.e. low values of both H$\beta$ and Fe$5335$ indices) corresponding to complex templates with a burst added over the secular SFH component.
This is clearly shown in the right panel of Fig.~\ref{fig:mgbhbeta}, where the average low scatter in metallicity at each bin of the index-index diagram confirms that we can effectively disentangle age and metallicity using this diagnostic plane.

%Since in the ultraviolet range we found that some indices are sensitive to the metallicity of the stellar populations, we explored the possibility to find a similar diagnostic plane in the UV. 
Figure~\ref{fig:2bl} shows, for the same set of galaxies, an index-index plane and its scatter based on the UV indices Fe$3619$ and BL$2538$. We can see that the metallicity values follow a roughly linear trend, where metallicity increases with increasing Fe$3619$, irrespective of the complexity of the SFHs. The small intrinsic scatter confirms the good quality of the diagnostic plane, showing a similar degeneracy as the (Fe$5335$,H$\beta$) one. Following the green arrows in Figs.~\ref{fig:mgbhbeta} and~\ref{fig:2bl}, again the effect is clear of a tiny fraction of a very young SSP over a $7$ Gyr one on the UV age-sensitive index BL$2538$, while the other index is barely affected.

%Notice that Fig.~\ref{fig:mgbhbeta} and~\ref{fig:2bl} show the behaviour of selected pairs of optical and UV indices in the models. 
However, given the nonzero observational errors (see white and green error bars in Fig.~\ref{fig:mgbhbeta} and~\ref{fig:2bl}) the uncertainties on inferring stellar metallicity using only two indices are not negligible. A better approach is to use the whole range of available spectral indices to optimally constrain the metallicity of stellar populations in galaxies, as we explore in the following sections.

\section{WEAVE-StePS-like observations}
\label{sec:stepsobs}
In this section, we explore the advantages offered by a wide spectral range provided by new-generation spectrographs, such as WEAVE and 4MOST. Since both spectrographs have similar performances in terms of resolution and wavelength coverage, we followed the same approach as in \cite{costantin2019few} by simulating observations of a large sample of galaxies as will be carried out in the WEAVE-StePS survey. From the library described in Section~\ref{sec:model}, we selected a statistically significant and representative chunk of $25000$ templates to produce mock WEAVE-StePS-like observations, while the remaining templates are used as comparison library. The selected chunk of templates adopted to reproduce mock observations contains twice the number that was used in \cite{costantin2019few} in order to have a better populated sample of old galaxies at different metallicities. Simulated spectra that realistically resemble the observations were obtained considering the throughput of the combined atmospheric transmission of the WHT and of the WEAVE spectrograph. The contributions to the noise are due to the extended Poisson noise from the sky background and to the read-out noise of the WEAVE CCDs.
In this way, our simulations  aptly reproduce a realistic S/N distribution as a function of wavelength for a variety of galaxy spectral types (including red galaxies), accounting for the reduced efficiency of the WEAVE+WHT system going to bluer wavelengths. The possible systematic effects present in real observed data (e.g. sky subtraction residuals, flux calibration errors) are not included, as their study is beyond the scope of the present paper.
For details on the procedure to transform the rest-frame subsample to mock WEAVE-StePS-like observations, we refer to Sect. $3$ in \cite{costantin2019few}. 

\begin{figure}
   \centering
   \includegraphics[width=0.5\textwidth]{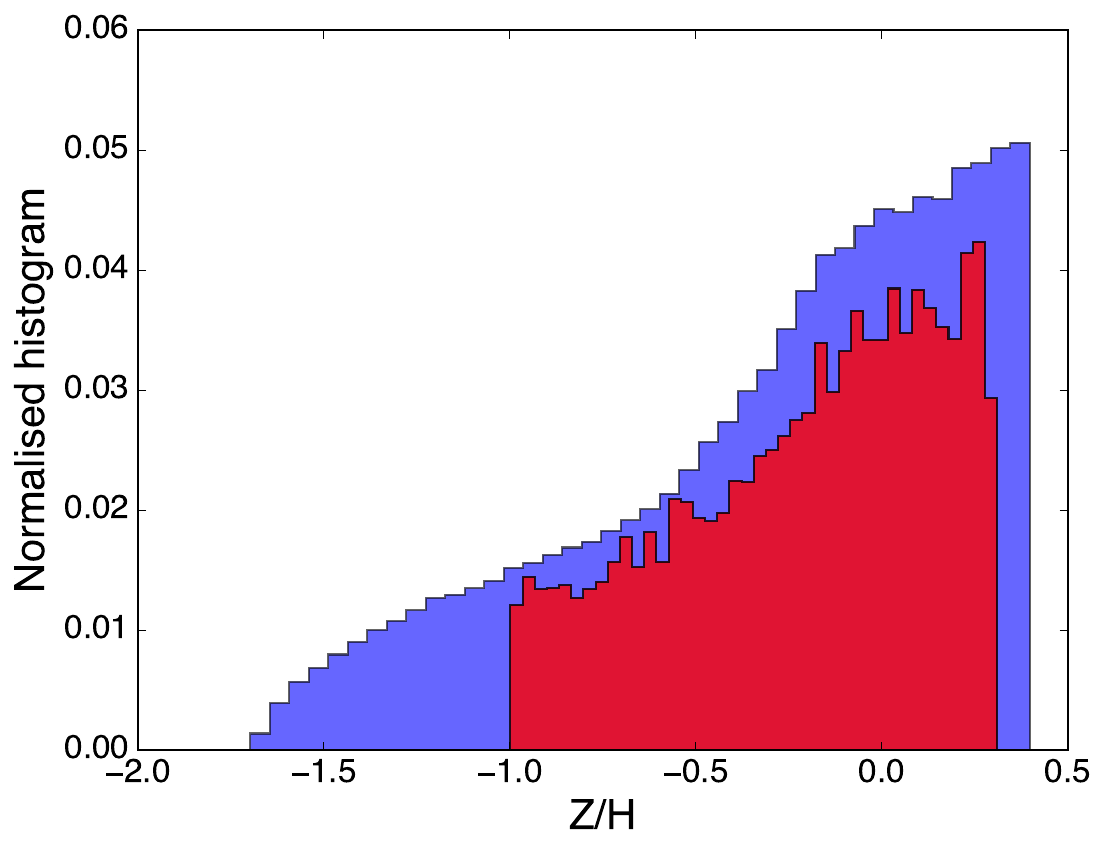}
      \caption{Distribution in metallicity of the mock WEAVE-StePS-like observations (red histogram) and the comparison library (blue histogram) at $z = 0.30$.
              }
         \label{fig:zh_distrib}
   \end{figure}

We considered three reference redshifts $z = [0.30, 0.55, 0.70]$ to reproduce the galaxies targeted by WEAVE-StePS and four bins of $S/N_{I,\textrm{obs}} = [10,20,30,50]\AA^{-1}$.
The $S/N_{I,\textrm{obs}} = 50\AA^{-1}$ bin has been added to the set adopted in \cite{costantin2019few} because, while not representing the expected WEAVE-StePS observations, it can provide more general useful insights on the possibility to estimate the stellar metallicity with higher S/N surveys, as it will be in 4MOST-StePS, or using a stacking procedure to increase the quality of observations at relatively low S/N.
For each redshift, observations have been created selecting from the set of $25000$ templates those corresponding to $t_{\textrm{form}} < \textrm{Age}_{\textrm{Universe}} -1$ Gyr at that redshift, where $t_{\textrm{form}}$ is the lookback time at the observation, $g$-band effective attenuation $A_{g} < 2$ mag, and mass-weighted metallicity $-1 < \mathrm{[Z/H]} < 0.3$. Figure~\ref{fig:zh_distrib} shows the distribution in metallicity of the mock WEAVE-StePS-like observations (red histogram) and the comparison library (blue histogram) at $z = 0.30$. Metallicities below $-1$ dex are extremely rare, in particular for massive galaxies as the targets of WEAVE-StePS survey. In fact, massive galaxies tend to have high metallicity values \citep[e.g.][]{gallazzi2005ages,thomas2010environment,gallazzi2021galaxy}. The final simulated sample consists of $9635$ spectra at $z = 0.30$, $8498$ at $z = 0.55,$ and $7884$ at $z = 0.70$, while the comparison library consists of more than $300000$ templates, selected to have $t_{\textrm{form}} < \textrm{Age}_{\textrm{Universe}}$. The selected subsamples are fully representative of the comparison library, without bias in physical properties, and large enough to explore the parameter space of the physical properties of the galaxies with robust statistics. Both the mock observations and the comparison library are convolved with a fixed velocity dispersion of $150$ kms$^{-1}$ in order to account for the typical velocity dispersion expected for WEAVE-StePS galaxies. The templates do not include emission lines, assuming that their contribution has been removed from the spectra.

  %----------------------------------------------------------------- 
%------------------

\section{Analysis}
\label{sec:analysis}
The main goal of our analysis is to exploit the information provided by key UV and optical absorption-line indices coupled with a Bayesian approach in order to infer the stellar metallicity of galaxies. The analysis is performed on the spectral data for each redshift and $S/N_{I,\textrm{obs}}$, as described in Section~\ref{sec:stepsobs}.

\subsection{Spectral indices}
\label{sec:spectralindices}
  %----------------------------------------------------------------- 
  % \begin{figure}
  % \centering
 %  \includegraphics[width=0.5\textwidth]{figures/threeindexsist_2.pdf}
 %     \caption{Difference between the measured values and the true ones for Fe$5335$, Fe$3619$ and Mgb, respectively, at $z = 0.30$ and $S/N_{I,\textrm{obs}} = 10$. The black solid lines represent the median values of each distribution. The filled region represents the $16-84$ confidence interval of the PDF.
 %             }
%         \label{fig:syst}
   %\end{figure}
%------------------
We considered a set of spectral indices to be used in our analysis, as listed in Table~\ref{tab:indices}. The indices we selected present some differences with respect to those reported in \cite{costantin2019few}. In fact, $12$ indices among the optical and UV ones have been removed, while 3 UV indices (BL$2720$, BL$2740$, BL$3619$) have been added.  Removed indices have been discarded because, even if they appear to be good metallicity or age indicators, they are strongly dependent on other unknown parameters (e.g. specific elements abundances or IMF) or they can be strongly affected by observational issues (e.g. accuracy of the relative flux calibration).
In particular, we excluded indices strongly sensitive to C, N, and Ti abundances (NH$3360$, CN$3883$, CN$4170$, Fe$4668$, Ca$4227,$ and Fe$5782$), because they could alter the final results, due to their strong dependence on abundance ratios. 
We also excluded indices potentially affected by chromospheric emission from the stellar atmospheres (MgII, CaHK) and those possibly affected by interstellar absorption (NaD). 
TiO$1$ and TiO$2$ have been excluded as they are very broad indices, possibly affected by uncertainty on flux calibration. Moreover, both TiO's are strongly dependent on the assumed IMF \citep[e.g.][]{spiniello2014stellar}. %Ca$4227$ has been excluded as it is very sensitive to individual abundance ratios (i.e. Ca/Fe and C/Fe) besides age and metallicity. 
Finally, we excluded the Mgwide because it is a very wide index that needs a reliable flux calibration of the observed data in the UV part of the spectral range, which is not easy to achieve in many spectroscopic surveys\footnote{We checked that adding the Mgwide in the inference of metallicity does not significantly change our main results, therefore we decided not to include this index in the main analysis.}.
In Table~\ref{tab:indices} we also show the variation of the UV indices when
adding $0.1\%$ of the total mass of very young SSP to an otherwise old SSP (see Sect.~\ref{sec:index}), normalised to the range of values covered by each index in the whole spectral library, namely, the relative sensitivity of each index to the presence of young populations. It is appreciable how the effect due to the presence of a tiny fraction of a very young population decreases as the wavelength increases, from $70\%$ of variation for the bluest UV index to $10\%$ for the reddest one.

We measured all the spectral indices that fall in the WEAVE observed spectral range at each selected redshift, and we estimated the observational errors for each index by generating $1000$ random gaussian distributed realisations within the noise of each single spectrum at any $S/N_{I,\textrm{obs}}$. For each realisation, we calculated the systematic and statistical errors as the median and the standard deviation of the relative difference between true and simulated values, respectively. Moreover, we added an extra error of $\sim 5\%$ to the D$_{\textrm{n}}4000$ to take into account the possible uncertainty of the spectrophotometric calibration that do not explicitly enter into our simulations, as in \cite{costantin2019few}. This extra error budget of $5\%$ has been estimated given the expected relative flux calibration errors for the future WEAVE-StePS survey within spectra windows of $200\AA$.
%Figure~\ref{fig:syst} shows the difference between the measured values and the true ones for Fe$5335$, Fe$3619$ and Mgb for all the simulated spectra at $z = 0.30$ and $S/N_{I,\textrm{obs}} = 10$. Even at the lowest $S/N_{I,\textrm{obs}}$, the median values are zero, meaning, as somehow expected, that there is no systematic error in measuring the indices.

\subsection{Bayesian inference}
The Bayesian approach provides a powerful framework for deriving the age and metallicity of stellar populations in galaxies. Our application of the Bayesian statistics consists on the comparison between a set of indices measured in the WEAVE-StePS-like simulations and those measured in the comparison library, as introduced by \cite{gallazzi2005ages} and also described in \cite{costantin2019few}. We computed the posterior probability of the age and metallicity parameters, where the likelihood is described by $\mathcal{L} = e^{-\chi^2/2}$, with

\begin{equation}
    \chi^2 = \sum_{i}{\left(\frac{I_{obs_i}-I_{mod_i}}{\sigma_{obs_i}}\right)}^2
,\end{equation}
where $I_{obs_i}$ and $I_{mod_i}$ are the $i$-th spectral index measured in the simulated and in the comparison spectra, respectively, and $\sigma_{obs_i}$ is the observational error of the $i$-th spectral simulated index. The observational error has been evaluated as the standard deviation of the distribution of $1000$ random gaussian realisations of the perturbed spectrum, as described in Section~\ref{sec:spectralindices}. %We assumed a logarithmically uniform prior in $r$-band light-weighted age and a uniform prior in metallicity (see Sect.~\ref{sec:model}).
With a Bayesian analysis we are able to retrieve for each observation the full probability density function (PDF) of any physical parameter, and we assume its median value as the expected value and the $16-84\%$ percentile range as the uncertainty of the estimated parameters. The PDF of a physical parameter for each simulated spectrum is given by the distribution in that parameter of the $\mathcal{L}$ of all the templates in the comparison library.

%%%%%%%%%%%%%%%%%%%%%%%%%%%%%%%%%%%%%%%%%%%%%%%%%%%%%%%%%%%%%%%%%%%%%%%
%%% TABLE

\begin{table}
\caption{UV and optical spectral indices.}
\centering
\begin{tabular}{ccccccc}
\hline\hline
UV index        &       \%      &       $z$             &       ref.             &   opt. index          &       $z$             &       ref.            \\
(1)             &       (1a)    &       (2)             &       (3)              &   (1)                 &       (2)             &       (3)             \\
\hline
FeII2402        &   70          &       > 0.66                  &        a                               &       D$_{\rm n}$4000         &    all                 &    d                          \\
BL2538      &    60    &   > 0.56               &    a                  &   H$\delta_{\rm F}$                     &        all                    &    e            \\
FeII2609        &    47    &   > 0.48          &    a                   &   H$\gamma_{\rm F}$   &  all                    &    e                  \\
BL2720      &      37       &   > 0.43          &    b              &   G$_{\rm band}$4300      &        all                    &    e                  \\
BL2740     & 44  & > 0.43               &    b                  &   Fe4383      &        all                     &    e                  \\
Mg2852          &   37          &   > 0.35              &    a                   &       Ca4455          &    all                &    e                   \\
Fe3000         &  23   &   > 0.31               &    a                  &   Fe4531                    &   all                     &    e                   \\
BL3096      &   23     &   > 0.25               &    a                  &   H$\beta$                  &      all                  &    e                   \\
BL3580     &     11    &   > 0.09               &    a                  &   Fe5015                &    all                &    e                  \\
Fe3619    &     10     &  > 0.06                        &    c                   &   Mgb              &    all                   &    e                   \\
               &        &                               &                       &            Fe5270             &    < 0.79                       &    e                   \\
          &             &                       &                       &   Fe5335                    &     < 0.77                &    e                   \\
                &                       &                       &                &   Fe5406                  &    < 0.75                 &    e                   \\
          &          &                  &                       &   Fe5709                  &     < 0.66                 &    e                  \\

\hline  
\end{tabular}
\tablefoot{(1) index name; (1a) variation of the UV index due to a $0.1\%$ young population on top of $99.9\%$ of $7$ Gyr population; (2) redshift range in which the index
is within the spectral range of the WEAVE spectrograph ($z_{\rm max} = 0.8$); (3) reference for 
indices definition: (a) \cite{fanelli1992spectral}, (b) \cite{chavez2007synthetic}, 
(c) \cite{gregg1994spectrophotometry}, (d) \cite{balogh1999differential} , (e) \cite{worthey1994old}.
}
\label{tab:indices}
\end{table}

%%%%%%%%%%%%%%%%%%%%%%%%%%%%%%%%%%%%%%%%%%%%%%%%%%%%%%%%%%%%%%%%%%%%%%%

\section{Metallicity estimates}
\label{sec:discussion}

  %----------------------------------------------------------------- 
   \begin{figure}
   \centering
   \includegraphics[width=0.5\textwidth]{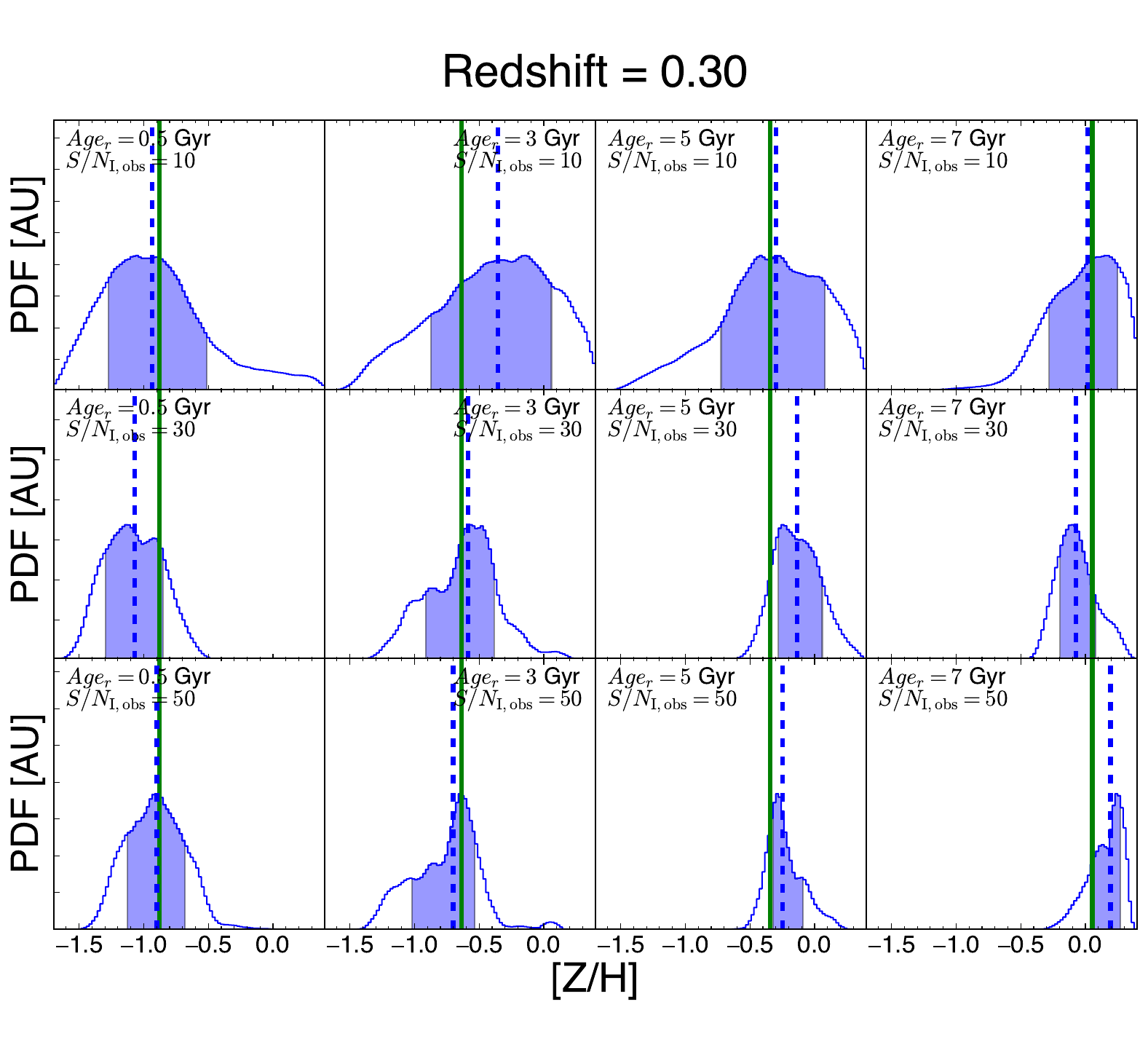}
      \caption{Examples of PDFs of mass-weighted metallicity of observations at $z = 0.3$ for $\text{Age}_r = [0.5,3,5,7]$ Gyr (from left to right) and $S/N_{I,\textrm{obs}} = [10,30,50]$ (from top to bottom). The filled region represents the $16-84$ confidence interval of the PDF. The solid green line marks the true value of metallicity, while the dashed blue line represents the median value of the corresponding distribution.
              }
         \label{fig:pdfzh}
   \end{figure}
%------------------
As detailed in the previous paragraph, we measured all optical and UV indices listed in Table~\ref{tab:indices} for the simulated and the comparison spectra ($17$ indices at $z = 0.3$, $22$ at $z = 0.55$ and $23$ at $z = 0.7$). Then we compared the observed spectral indices with those obtained in the comparison library taking into account the expected observational errors to finally obtain metallicity measurements of our WEAVE-StePS-like observations by marginalising over the other parameters.

Figure~\ref{fig:pdfzh} shows examples of metallicity PDFs of four simulated galaxies which differ for the mean $r$-band light-weighted ages of their stellar content ($\text{Age}_r = [0.5,3,5,7]$ Gyr), and for the mass-weighted metallicity ([Z/H] = [$-0.25,-0.9,0.5,0.2$] dex) and $S/N_{I,\textrm{obs}} = [10,30,50]$, respectively. It is noticeable that the uncertainty on the metallicity estimates decreases with increasing age, in particular at low $S/N_{I,\textrm{obs}}$. Moreover, $S/N_{I,\textrm{obs}}$ plays a crucial role in the uncertainty of the derived metallicity estimate, and differences in the PDF width between $S/N_{I,\textrm{obs}} = 10$ and $S/N_{I,\textrm{obs}} = 30$ are quite large. However, it clearly appears that from $S/N_{I,\textrm{obs}} = 30$ to $S/N_{I,\textrm{obs}} = 50$ the improvement in the uncertainty of the metallicity measurements is smaller, in particular for old ages.

  %----------------------------------------------------------------- 
   \begin{figure}
   \centering
   \includegraphics[width=0.5\textwidth]{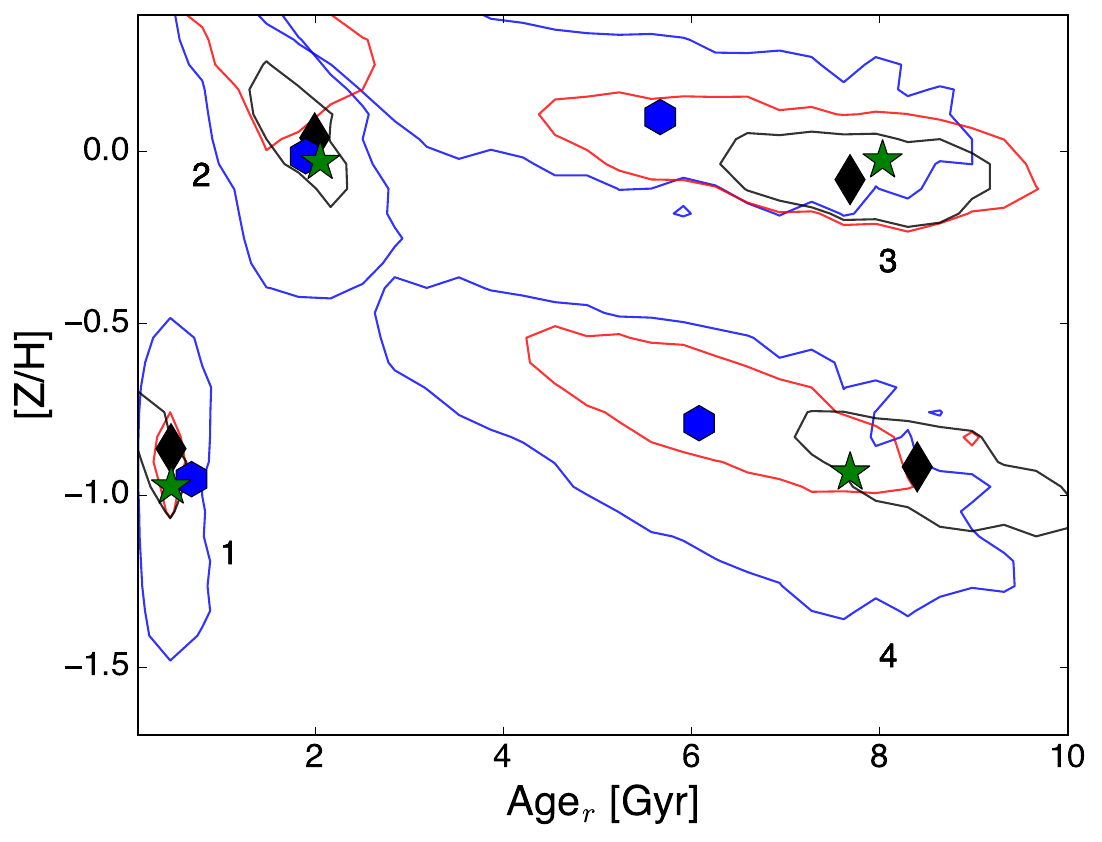}
      \caption{Examples of $68\%$ confidence interval of joint $r$-band light-weighted age and mass-weighted metallicity distributions at $z = 0.3$ at $S/N_{I,\textrm{obs}} = 10$ (blue lines), $S/N_{I,\textrm{obs}} = 30$ (red lines), and $S/N_{I,\textrm{obs}} = 50$ (black lines). The  blue hexagons marks the median age-metallicity values of the distribution at $S/N_{I,\textrm{obs}} = 10$, the black diamonds the median age-metallicity at $S/N_{I,\textrm{obs}} = 50$, and the green stars represent the true age-metallicity values.
              }
         \label{fig:pdf2d}
   \end{figure}
%------------------

Figure~\ref{fig:pdf2d} shows examples of joint age and metallicity probability distributions for another four simulated galaxies with different values of age and metallicity at $z = 0.3$, where the contours with different colours (blue, red, and black) correspond to different $S/N_{I,\textrm{obs}} = [10,30,50]$, respectively. The age-metallicity degeneracy can be clearly seen for galaxy n.$2$ and $4$, for which the corresponding probability distributions are elongated diagonally in the diagram, in particular at $S/N_{I,\textrm{obs}} = 10$. It can also be seen that the constraints on the metallicity estimates are tighter with increasing age.  
This is expected, as already noticed in Sect.~\ref{sec:index}, because for old stellar populations metallic lines become stronger and their variation becomes steeper and independent of age, giving better constraints. Instead, age indicators (e.g.  Balmer lines) become weaker and their variation flatter for older populations. On the contrary, metallicity indicators (both optical and UV) become weaker for hot stars, and change very little with metallicity at ages lower than $2$ Gyr, becoming mainly sensitive to the age.

   \begin{figure*}
   \centering
   \subfloat[][\emph{}]
        {\includegraphics[width=0.95\textwidth]{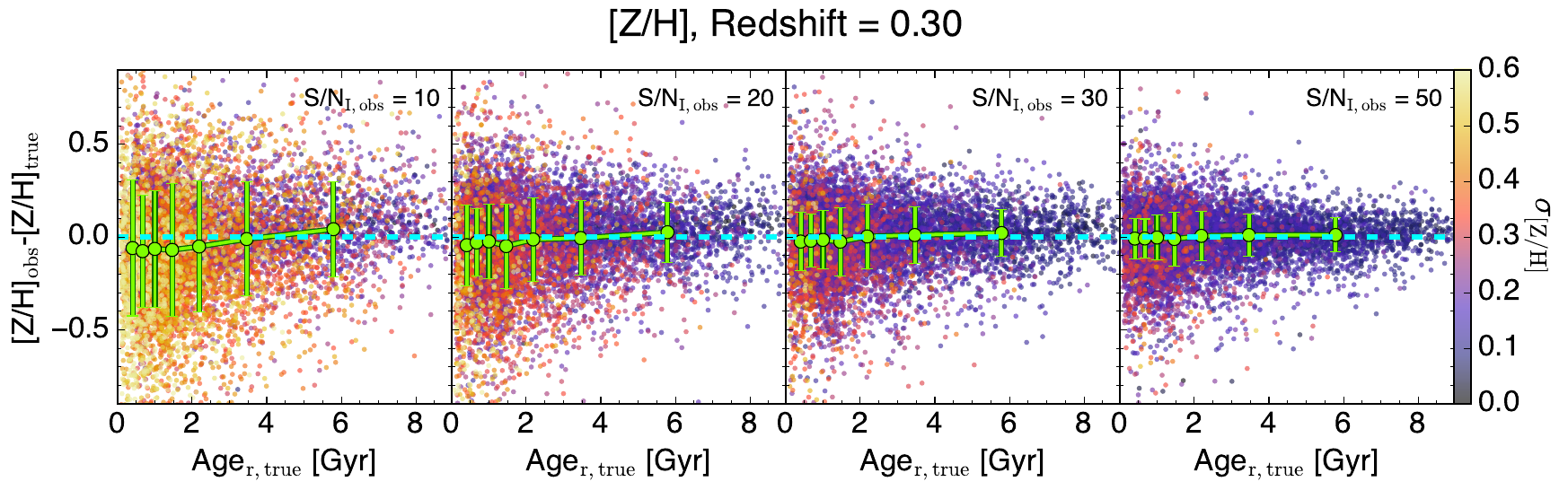}} \quad
        \subfloat[][\emph{}]
        {\includegraphics[width=0.95\textwidth]{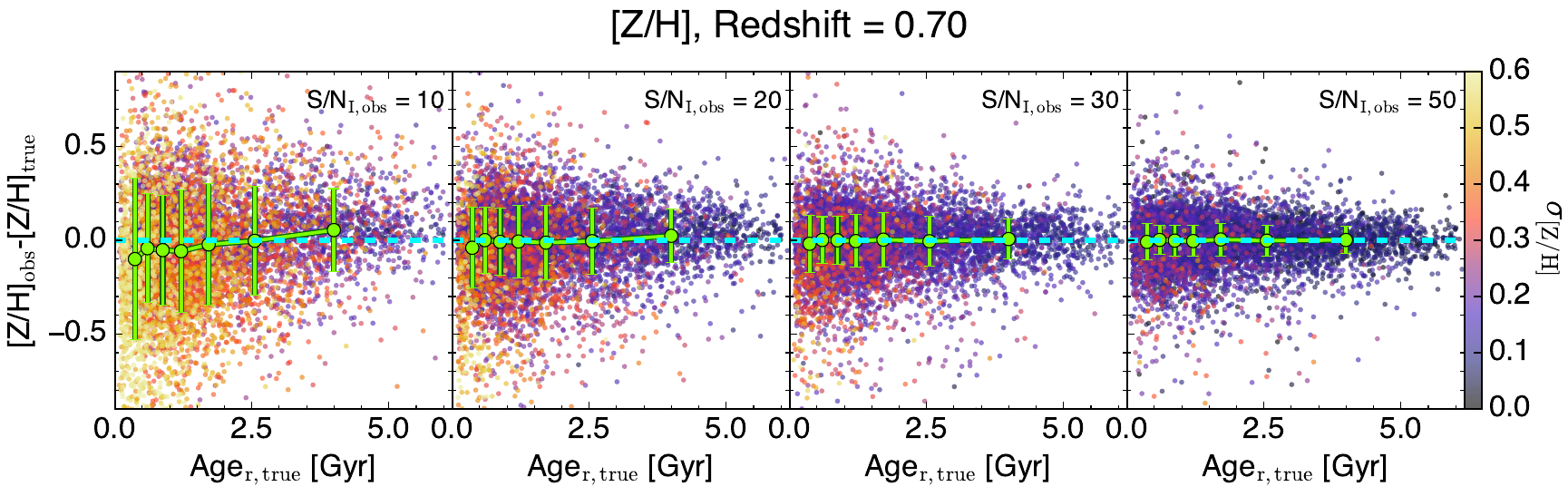}} \quad
 
   \caption{Difference between observed and true metallicity, as a function of true $r$-band light-weighted age values at different redshifts. \textit{Top panel}: Difference between observed and true metallicity, as a function of true $r$-band light-weighted age values at $z = 0.30$. Each dot is colour-coded according to $\sigma_\textrm{[Z/H]}$ obtained from the posterior. The green dots are the median values of the [Z/H] differences in bins of Age$_{r,\text{true}}$, while the error bars are the median of the $\sigma_\textrm{[Z/H]}$, with each bin having the same number of data points. The cyan dashed line represents the $0$ value of the y-axis. \textit{Bottom panel}: Same as the top panel for $z = 0.70$.}
              \label{fig:scatterplotdn40000}
    \end{figure*}

\begin{table*}
\caption{Error estimates of all the simulated galaxies reported in seven $r$-band light weighted age bins at all the redshift and $S/N_{I,\textrm{obs}}$. The total number of available galaxies has been divided in seven bins equally populated. Each bin has different width, assuming the median value as reference. The errors are the $68\%$ confidence interval of the PDF of mass-weighted stellar metallicity, and are expressed in dex.}             % title of Table
\label{table:summaryerrorsage}      % is used to refer this table in the text
\centering                          % used for centering table
\begin{tabular}{l l l l l l l l}        % centered columns (4 columns)
\hline\hline                 % inserts double horizontal lines
\multicolumn{8}{c}{$z = $ 0.30}   \\    % table heading 

Age$_{bin}$ [Gyr] & [0.07, 0.5] & [0.5, 0.8] & [0.8, 1.1] & [1.1, 1.7] & [1.7, 2.7] & [2.7, 4.4] & [4.4, 9]\\
\hline                        % inserts single horizontal line

   $S/N_{I,\textrm{obs}} = 10$ &0.37 & 0.30 & 0.32 & 0.36 & 0.35 & 0.30& 0.26\\      % inserting body of the table
   $S/N_{I,\textrm{obs}} = 20$ & 0.22 & 0.18 & 0.20 & 0.23 &0.22 & 0.20 & 0.16\\

   $S/N_{I,\textrm{obs}} = 30$ & 0.16 & 0.14 & 0.16 & 0.18 & 0.18 & 0.15 & 0.12\\
   
   $S/N_{I,\textrm{obs}} = 50$ & 0.11 & 0.11 & 0.12 & 0.14 & 0.13 & 0.11 & 0.09\\
   \hline\hline

    \multicolumn{8}{c}{$z = $ 0.55}  \\ 
    Age$_{bin}$ [Gyr] & [0.07, 0.5] & [0.5, 0.7] & [0.7, 1]  & [1, 1.4] & [1.4, 2.2] & [2.2, 3.5] &  [3.5, 6.8] \\ 
     \hline
 
   $S/N_{I,\textrm{obs}} = 10$ & 0.39 & 0.29 & 0.30 &0.34 & 0.33 & 0.29& 0.23 \\ 
   $S/N_{I,\textrm{obs}} = 20$ & 0.22 & 0.18 & 0.18 & 0.21 & 0.21 & 0.19 & 0.15 \\
   $S/N_{I,\textrm{obs}} = 30$ & 0.16 & 0.13 & 0.15 & 0.17 & 0.17 & 0.15 & 0.12\\

   $S/N_{I,\textrm{obs}} = 50$ & 0.10 & 0.09 & 0.11 & 0.12 & 0.11 & 0.10 &0.08\\
      \hline\hline

    \multicolumn{8}{c}{$z = $ 0.70} \\ 
    Age$_{bin}$ [Gyr] & [0.07, 0.5] & [0.5, 0.7] & [0.7, 0.9]  & [0.9, 1.3] & [1.3, 2]  & [2, 3.1] & [3.1, 6.1] \\ 
     \hline
   $S/N_{I,\textrm{obs}} = 10$ & 0.43 & 0.29 &0.29 & 0.32 & 0.33 & 0.29 & 0.22\\

   $S/N_{I,\textrm{obs}} = 20$ & 0.22 & 0.18 & 0.18 & 0.19 & 0.20 & 0.18 & 0.14\\
 
   $S/N_{I,\textrm{obs}} = 30$ & 0.15 & 0.12 & 0.14 & 0.14 & 0.14 & 0.13 & 0.11 \\

   $S/N_{I,\textrm{obs}} = 50$ & 0.10 & 0.08 & 0.08 & 0.08 & 0.09 & 0.09 &0.07 \\
   
\hline                                   %inserts single line
\end{tabular}
\end{table*} 
   
\begin{figure*}
   \centering
   \subfloat[][\emph{}]
        {\includegraphics[width=0.95\textwidth]{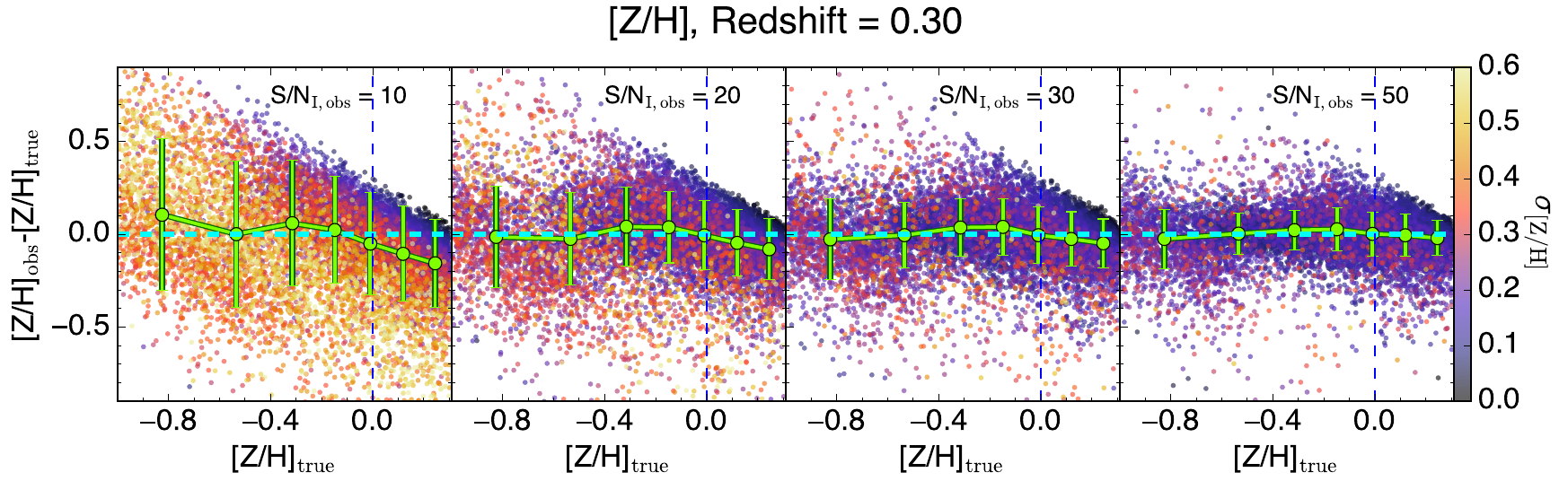}} \quad
        \subfloat[][\emph{}]
        {\includegraphics[width=0.95\textwidth]{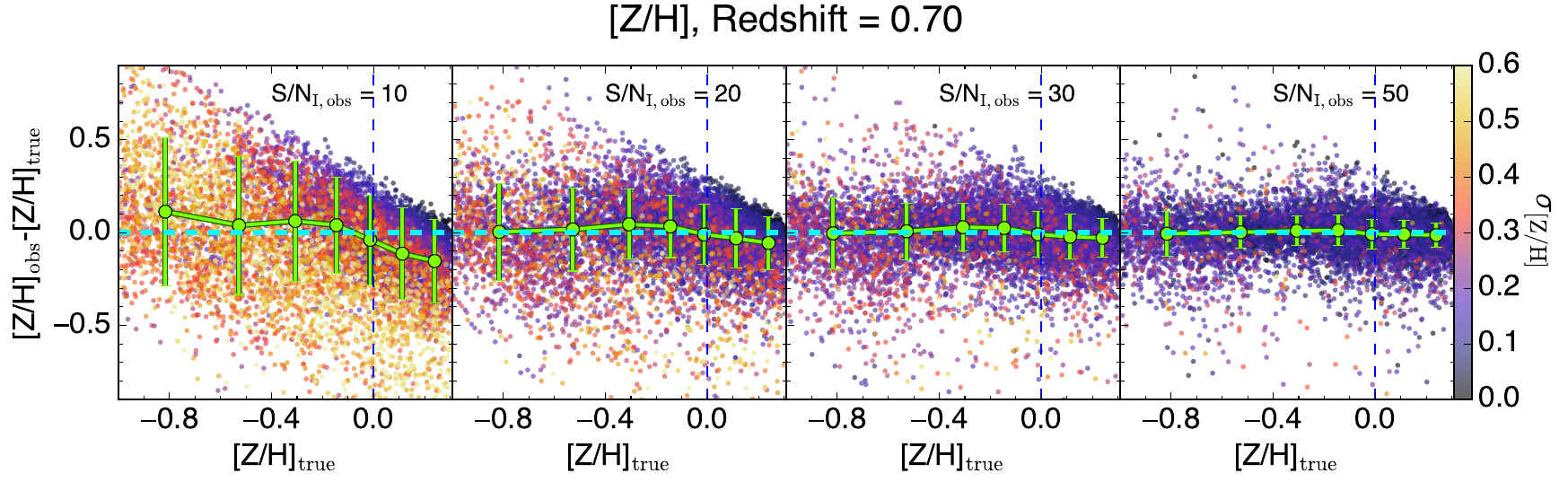}} \quad
 
   \caption{Difference between observed and true metallicity as a function of the true values of metallicity at different redshifts. \textit{Top panel}: Difference between observed and true metallicity as a function of the true values of metallicity at $z = 0.30$. Each dot is colour-coded according to $\sigma_\textrm{[Z/H]}$ obtained from the PDF. The green dots are the median values of the [Z/H] differences in bins of Age$_{r,\text{true}}$, while the error bars are the median of the $\sigma_\textrm{[Z/H]}$, with each bin having the same number of data points. The cyan dashed line represents the $0$ value of the y-axis. \textit{Bottom panel}: Same as the top panel for $z = 0.70$.}
              \label{fig:scatterplotxzh}
    \end{figure*}

Figure~\ref{fig:scatterplotdn40000} shows the difference between the observed metallicity and the true one as a function of the $r$-band light-weighted age for the simulations at $z = [0.3,0.7]$ and $S/N_{I,\textrm{obs}} = [10,20,30,50]$. Figure~\ref{fig:scatterplotxzh} shows the same difference but as a function of the true metallicity values. Points are colour-coded according to the metallicity errors obtained from the PDF.
Table~\ref{table:summaryerrorsage} summarises our capability to retrieve the mass-weighted metallicity (as presented in Fig.~\ref{fig:scatterplotdn40000}) by showing the $1\sigma$ errors for seven age bins with equal number of data points, at all the simulated redshift and $S/N_{I,\textrm{obs}}$, while Table~\ref{table:summaryerrorsmet} shows the same $1\sigma$ errors for seven metallicity bins. In this table, we also added the errors estimated at $S/N_{I,\textrm{obs}} = 10$ for simulated galaxies with $r$-band light-weighted age $> 2$ Gyr, showing that for older populations, we obtain lower uncertainties on metallicity. It is worth noticing that the typical uncertainties of metallicity estimates well match the median dispersion around the true values.

\begin{table*}
\caption{Same as Table~\ref{table:summaryerrorsage} with simulated galaxies reported in seven metallicity bins. The values in parenthesis represent the error estimates at $S/N_{I,\textrm{obs}} = 10$ for simulated galaxies with $r$-band light-weighted age $> 2$ Gyr.}             % title of Table
\label{table:summaryerrorsmet}      % is used to refer this table in the text
\centering                          % used for centering table
\begin{tabular}{l l l l l l l l}        % centered columns (4 columns)
\hline\hline                 % inserts double horizontal lines
 \multicolumn{8}{c}{$z = $ 0.30} \\    % table heading 

[Z/H]$_{bin}$ [dex] & [$-$1, $-$0.67] & [$-$0.67, $-$0.42] & [$-$0.42, $-$0.22] & [$-$0.22, $-$0.06]& [$-$0.06, 0.09] & [0.09, 0.21] & [0.21, 0.31]\\
\hline                        % inserts single horizontal line

   $S/N_{I,\textrm{obs}} = 10$ & 0.42 (0.39) & 0.39 (0.39) & 0.33 (0.33) & 0.28 (0.28) & 0.26 (0.25) & 0.25 (0.21)& 0.23 (0.19)\\      % inserting body of the table
   $S/N_{I,\textrm{obs}} = 20$ & 0.28 & 0.24 & 0.20 &0.18 & 0.18 & 0.17 & 0.16\\

   $S/N_{I,\textrm{obs}} = 30$ & 0.22 & 0.17 & 0.15 & 0.15 & 0.14 &0.14 & 0.12\\
   
   $S/N_{I,\textrm{obs}} = 50$ &0.16 & 0.10 & 0.10 & 0.11 & 0.11 & 0.11 & 0.09\\
   \hline\hline
    \multicolumn{8}{c}{$z = $ 0.55}  \\ 
    $\text{[Z/H]}_{bin}$ [dex] & [$-$1, $-$0.67] & [$-$0.67, $-$0.42] & [$-$0.42, $-$0.22] & [$-$0.22, $-$0.06]& [$-$0.06, 0.09] & [0.09, 0.21] & [0.21, 0.31]\\ 
     \hline
 
   $S/N_{I,\textrm{obs}} = 10$ & 0.40 (0.36) &0.37 (0.36) & 0.31 (0.31) & 0.26 (0.26) & 0.25 (0.22) & 0.24 (0.19)& 0.22 (0.17) \\ 
   $S/N_{I,\textrm{obs}} = 20$ & 0.26 & 0.22 & 0.19 & 0.17 & 0.17 & 0.16 & 0.15 \\
   $S/N_{I,\textrm{obs}} = 30$ & 0.21 & 0.16 & 0.14 & 0.14 & 0.13 & 0.13 & 0.12\\

   $S/N_{I,\textrm{obs}} = 50$ & 0.14 & 0.09 & 0.09 & 0.09 & 0.09 & 0.09 &0.08\\
      \hline\hline
    \multicolumn{8}{c}{$z = $ 0.70}  \\ 
    $\text{[Z/H]}_{bin}$ [dex] & [$-$1, $-$0.67] & [$-$0.67, $-$0.42] & [$-$0.42, $-$0.22] & [$-$0.22, $-$0.06]& [$-$0.06, 0.09] & [0.09, 0.21] & [0.21, 0.31] \\ 
     \hline
   $S/N_{I,\textrm{obs}} = 10$ & 0.41 (0.36) & 0.37 (0.34) & 0.31 (0.29) & 0.25 (0.25) & 0.24 (0.22) & 0.24 (0.18) & 0.22 (0.16)\\

   $S/N_{I,\textrm{obs}} = 20$ & 0.27 & 0.22 &0.18 & 0.16 & 0.15 & 0.15 & 0.13\\
 
   $S/N_{I,\textrm{obs}} = 30$ & 0.19 & 0.15 & 0.13 & 0.12 & 0.12 & 0.11 & 0.10 \\

   $S/N_{I,\textrm{obs}} = 50$ & 0.12 & 0.08& 0.08 & 0.08 & 0.08 & 0.07 &0.06 \\
   
\hline                                   %inserts single line
\end{tabular}
\end{table*}

The main result from Fig.~\ref{fig:scatterplotdn40000} is that the median differences between the measured and the true metallicities as a function of the $r$-band light-weighted age are consistent with $0$ for all the cases explored. This indicates that, on average, metallicity can be inferred with no systematic deviations at any galaxy age for all the $S/N_{I,\textrm{obs}}$. On the other hand, the typical median errors decrease with increasing $r$-band age for all the $S/N_{I,\textrm{obs}}$, with values, at $S/N_{I,\textrm{obs}} = 10$, from $\sigma = 0.38$ dex for $\text{age}_{\text{bin}} = [0.07, 0.5]$ Gyr to $\sigma = 0.26$ dex for $\text{age}_{\text{bin}} = [4.4, 9]$ Gyr, improving $\sim 0.1$ dex at older ages. A similar effect is obtained by increasing the $S/N_{I,\textrm{obs}}$ from $10$ to $20$, reducing the uncertainties for older galaxies down to less than $0.2$ dex.
The lower panels of Figure~\ref{fig:scatterplotdn40000} show the results for simulations at $z=0.70$. As the number and the typical errors of available indices falling into the WEAVE wavelength range varies with the redshift considered, the upper and lower panels of Figure~\ref{fig:scatterplotdn40000} differ. Errors in the metallicity estimate at $z=0.7$ for galaxies older than $1$ Gyr are slightly lower than those obtained at $z=0.3$, up to $0.04$ dex around $5$ Gyr. This is mainly due to the higher number of UV indices available at higher redshift.

  %----------------------------------------------------------------- 
   \begin{figure}
   \centering
   \includegraphics[width=0.5\textwidth]{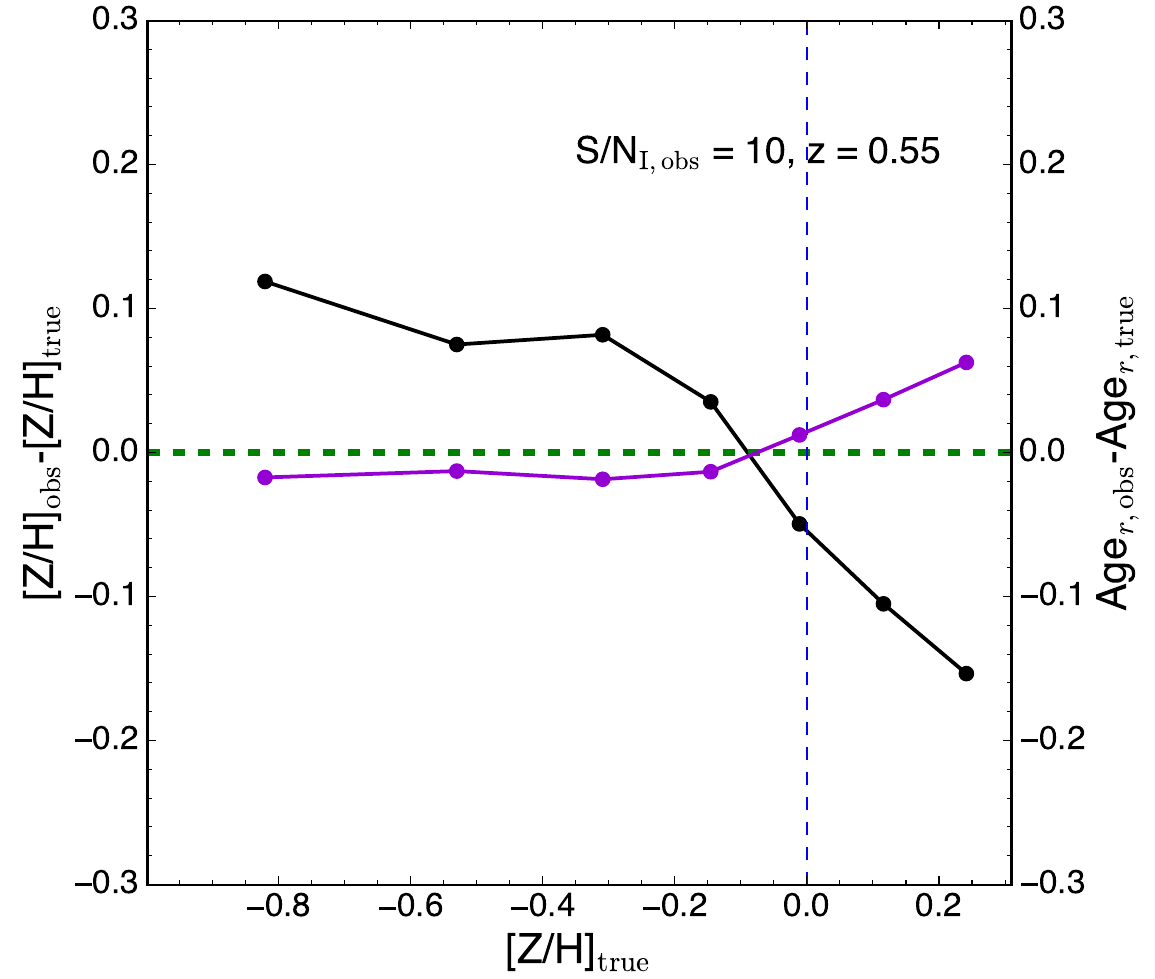}
      \caption{Median values of the difference between observed and true values of mass-weighted metallicity (black points and lines) and $r$-band light-weighted age (violet point and lines), as a function of true metallicity at $z = 0.55$ and $S/N_{I,\textrm{obs}} = 10$. The blue vertical dashed line marks the solar metallicity value ([Z/H] $= 0$). The green dashed line represents the $0$ value of the y-axis.
              }
         \label{fig:confrzhage}
   \end{figure}
%------------------

The expected precision in the metallicity estimate as a function of the true metallicity values is shown in Fig.~\ref{fig:scatterplotxzh}, both at low (upper panel) and at high (lower panel) redshift. At $S/N_{I,\textrm{obs}} = 10$, the uncertainties decrease from $\sigma = 0.37$ dex for very low metallicity (i.e. [Z/H] $ \le -0.8$) to $\sigma = 0.22$ dex for super-solar metallicities. Concerning the expected accuracy, at the lowest/highest metallicities, there is a overestimation or underestimation of $\sim 0.1$ dex. To better appreciate this trend, Figure~\ref{fig:confrzhage} shows the median values in seven metallicity bins of the difference between the observed and true values of mass-weighted metallicity compared to the difference for $r$-band light-weighted age as a function of the true values of metallicity at $z = 0.55$ and $S/N_{I,\textrm{obs}} = 10$. When metallicity is underestimated, age is overestimated -- and the other way around as well, consistently with the well-known age-metallicity degeneracy. Moreover the offset at the highest metallicity likely reflects the lack of templates above [Z/H] $\sim0.3$ (the maximum value in the input library), making the lowest [Z/H] the most preferable in the index analysis (see Sect.~\ref{sec:spectralindices}). In other words, at the highest as well as the lowest metallicity, our metallicity estimates, for $S/N_{I,\textrm{obs}} = 10$, are affected by the prior distribution. Nevertheless, the systematic offset is well within the errors ($\sim 0.3$ dex at $S/N_{I,\textrm{obs}} = 10$), and it drastically decreases already at $S/N_{I,\textrm{obs}} = 20$. 

  %----------------------------------------------------------------- 
   \begin{figure}
   \centering
   \includegraphics[width=0.5\textwidth]{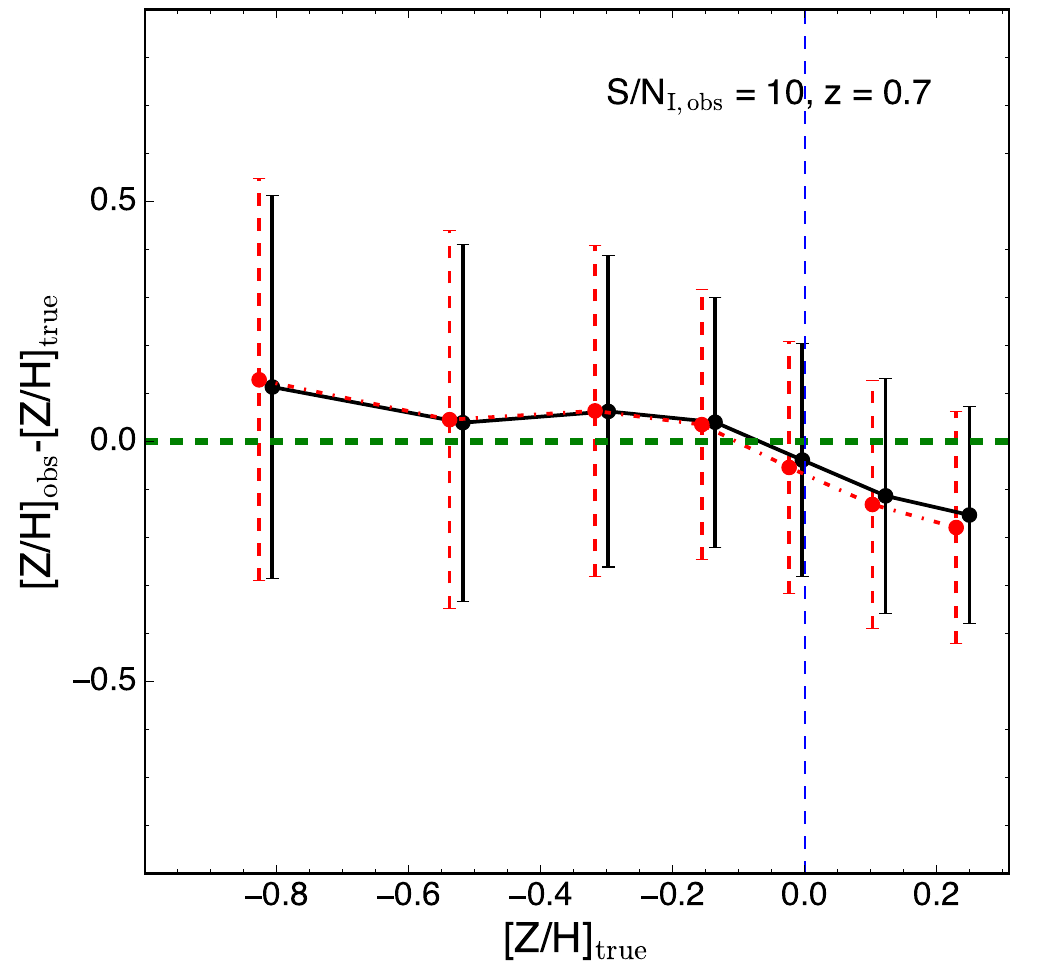}
      \caption{Median values of the difference between observed and true metallicity as a function of the true values of metallicity at $z = 0.70$ and $S/N_{I,\textrm{obs}} = 10$. The black error bars indicates the main case of our work, while the red dashed error bars represent the case excluding indices below $3000\AA$. The blue vertical dashed line marks the solar metallicity value ([Z/H] $=0$). The green dashed line represents the $0$ value of the y-axis.
              }
         \label{fig:confrno3000}
   \end{figure}
%------------------

The results displayed in Figs.~\ref{fig:scatterplotdn40000},~\ref{fig:scatterplotxzh}, and~\ref{fig:confrzhage} are based on all the available indices listed in Table~\ref{tab:indices} depending on the simulated redshift. However, as already noted in Sect.~\ref{sec:index}, UV indices located at restframe wavelength lower than $3000\AA$ are strongly affected by the presence of even a tiny fraction of hot stars (e.g. very young or PAGB/HB stars). 
Since both the origin and the effect of these populations are still poorly understood and hard to 
model \citep[see e.g.][]{le2016modelling,salvador2020sub,salvador2021young,salvador2022lessons}, and
an extensive treatment of these components is beyond the scope of the present paper,
we also repeated our analysis by excluding UV indices at $\lambda < 3000\AA$ (i.e. the ones potentially affected by these components, see Table~\ref{tab:indices}). Figure~\ref{fig:confrno3000} shows the comparison between the main analysis of this work and the same analysis performed excluding UV indices below $3000\AA$. We found very similar uncertainties in both cases, thus, our ability to constrain metallicity is the same when excluding UV indices below $3000\AA$. 
We also verified that the age limit within the comparison library, selected to be lower than the age of the universe at each redshift, does not affect the results, which are consistently derived even using the entire available range of ages.
We checked the effect on the results of simulating galaxies based on $\alpha$-enhanced stellar populations (which is particularly relevant for massive galaxies) keeping the comparison template library based on solar chemical composition (see Appendix~\ref{sec:appA} for details). Although we found that the effect of [$\alpha$/Fe] $= 0.4$ dex can introduce systematic offsets in metallicity estimates up to $-0.3$ dex at $S/N = 10$, our test is an extreme case, since usually massive quiescent galaxies have [$\alpha$/Fe] up to $0.3$ dex \citep[e.g.][]{thomas2010environment,la2013spider,carnall2022stellar}. Moreover, the effect of varying chemical abundances might be not so
severe, as the overabundance of different elements affect spectral features in a different way. For instance, the relative response of an SSP model spectrum to increasing [$\alpha$/Fe] tends to go into the opposite direction of increasing [C/Fe], so that for several spectral indices, increasing both elements (as it is the case in massive galaxies) affects only marginally the line-strengths \citep[see][]{la2017imf}. In future works, we will take this issue into account by using models with varying [$\alpha$/Fe] and other elemental abundances.

  %----------------------------------------------------------------- 
   \begin{figure}
   \centering
   \includegraphics[width=0.5\textwidth]{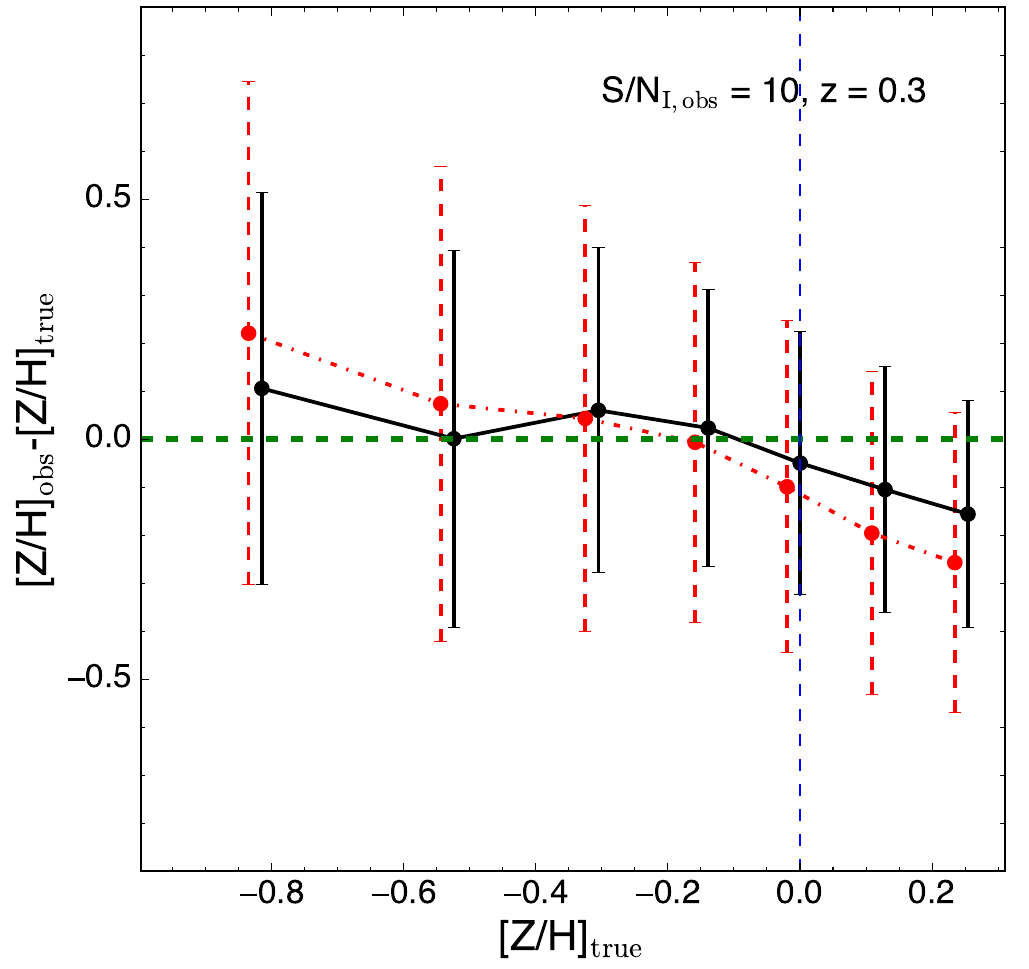}
      \caption{Median values of the difference between observed and true metallicity as a function of true metallicity values at $z = 0.30$ and $S/N_{I,\textrm{obs}} = 10$. The black error bars show results for our analysis including all spectral indices, while the red dashed error bars represent the case using only the $5$ absorption indices used in \cite{gallazzi2005ages}. The blue vertical dashed line marks the solar metallicity value ([Z/H] $= 0$). The green dashed line represents the $0$ value of the y-axis.
              }
         \label{fig:confrgal}
   \end{figure}
%------------------

We compared the accuracy of our metallicity estimates for simulated WEAVE-StePS-like spectra with those reported in the literature.
Using SDSS spectra, \cite{gallazzi2005ages} inferred metallicity of early-type and late-type galaxies in the Local Universe using five absorption indices, of which three  are composites (D$_n4000$, H$\delta_A+$H$\gamma_A$, H$\beta$, [Mg$_2$Fe], [MgFe]').
%Selecting a sample similar to that of \cite{gallazzi2005ages}, considering their D$_n4000$ and H$\delta_A+$H$\gamma_A$ indices bins at $z = 0.30$, we find that at $S/N_{I,\textrm{obs}} = 10$ the errors expected from our work based on more indices are lower than those reported by \cite{gallazzi2005ages} up to $0.2$ dex.  In particular, we found $\sigma_{\text{our work}} = [0.35,0.31,0.18,0.16]$ dex, compared to $\sigma_{Gallazzi} = [0.57, 0.43, 0.27, 0.12]$.
To test the advantage of using a large number of indices, we performed our analysis considering only the five indices adopted by \cite{gallazzi2005ages} and compare the results with our more general simulations including all the available indices at redshift $z = 0.3$. Both analyses are based on S/N derived from WEAVE-StePS-like simulations, and the results are presented in Fig.~\ref{fig:confrgal}. 
Both the systematic offset between true and measured values, and the errors affecting the metallicity estimates are lower (up to $0.1$ dex) when using all the available indices with respect to the restricted number of indices in \cite{gallazzi2005ages}. 

Another instructive comparison can be made with the work of \cite{choi2014assembly}, who inferred the metallicity by performing full spectral fittings in the $4700-5500\AA$ range on simulated spectra \citep[see Fig. A.1 of][]{choi2014assembly}. They considered mock observations for a sample of quiescent galaxies at age $\sim 6$ Gyr and solar metallicity, and obtained metallicity values without particular systematic offset and with error $\sigma_{\text{Choi}} \sim 0.15$ dex. Considering a sample of galaxies from our simulations in the same age and metallicity range as in \cite{choi2014assembly}, we found an error of $0.15$ dex in [Z/H], perfectly matching the result obtained in \cite{choi2014assembly}. This result is mainly due to the large set of indices used in our analysis.

\cite{lopez2016simultaneous} performed a full spectral fitting analysis in the $3700-6800\AA$ range on simulated CALIFA \citep[Calar Alto Legacy Integral Field Area,][]{sanchez2012califa} spectra to infer stellar metallicity. They considered mock observations with age and metallicity parameters similar to those of the sample used in the analysis of this work, and they obtained metallicity values without particular systematic offset from the true value and with error $\sigma_{\text{Lopez}}$ = [$0.33, 0.22, 0.13$] at S/N = [10, 20, 50]. Considering the metallicity constraints on our entire sample at $z = 0.30$, we found errors of [$0.32,0.21,0.11$] at $S/N = [10,20,50]$, consistent to the results obtained in \cite{lopez2016simultaneous}.
Spectral fitting is expected to provide lower error bars, as   all the information in the galaxy spectrum is used. Nevertheless, spectral indices help to extract the cleanest information from a galaxy spectrum, through well-selected and characterised spectral features, and help to reduce the effect of other parameters on the estimates of the metallicity (e.g. presence of dust and possible uncertainties on flux calibration). Moreover, as it will be explained in Sect.~\ref{sec:bluer}, increasing the number of indices helps to decrease the errors in constraining the metallicity. 

\section{Constraints from bluer indices}
\label{sec:bluer}
  %----------------------------------------------------------------- 
   \begin{figure*}
   \centering
   \includegraphics[width=0.95\textwidth]{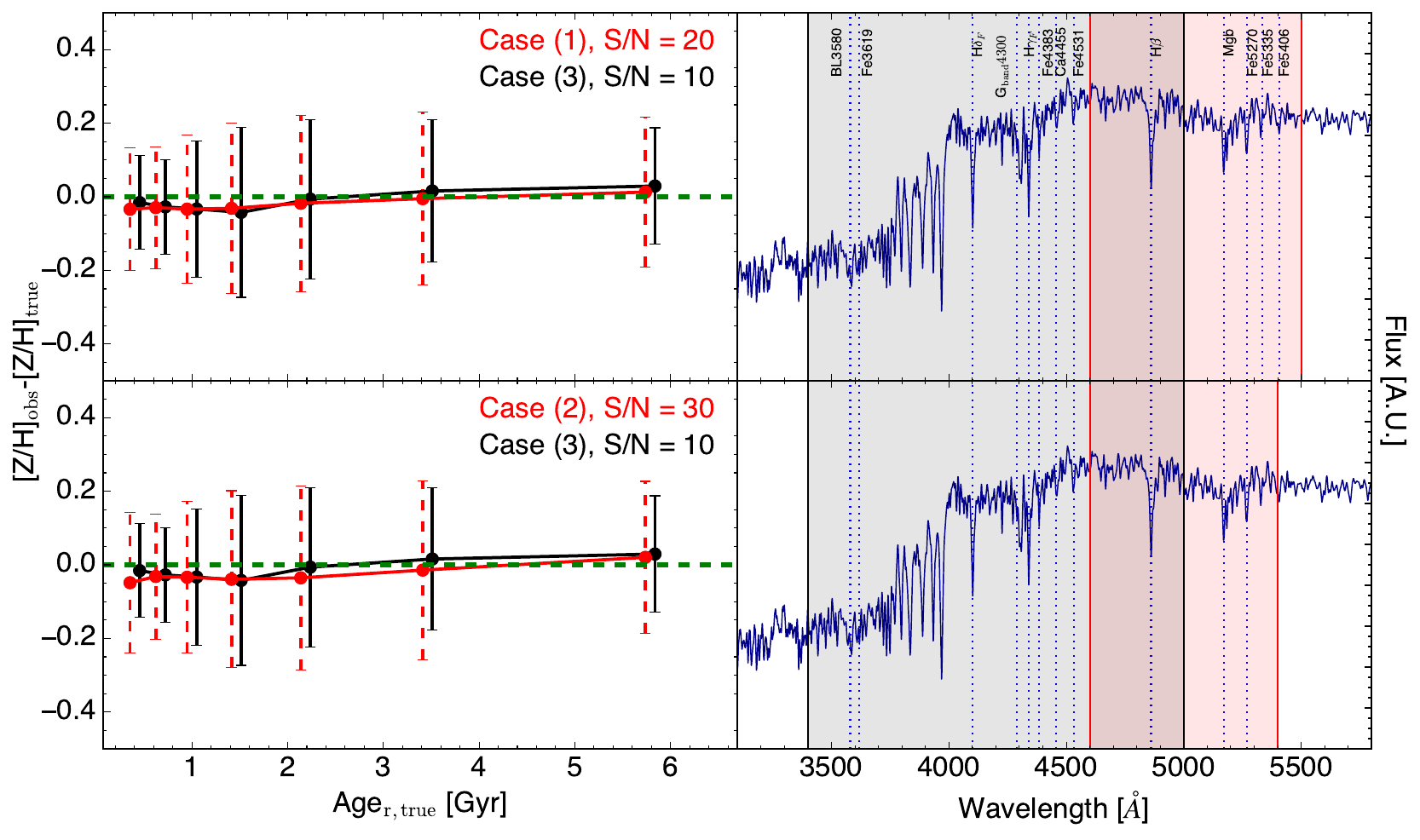}
  \caption{Median values of the difference between observed and true metallicity as a function of true $r$-band light-weighted age values. In both left panels, black error bars denotes case $3$ ([BL$3580$, Fe$3619$, H$\delta_\text{F}$, H$\gamma_\text{F}$, G$_\text{band}4300$, Fe$4383$, Ca$4455$, Fe$4531$, H$\beta$]) at S/N = 10, while the red dashed error bars indicates case $1$ ([H$\beta$, Mgb, Fe$5270$, Fe$5335$, Fe$5406$]) at S/N = 20 (upper left panel) and case $2$ ([H$\beta$, Mgb, Fe$5270$]) at S/N = 30 (lower left panel). The green dashed line in the left panels represents the $0$ value of the y-axis. The blue curves in the right panels represent an example of a rest-frame template with $3$ Gyr and solar metallicity. The black filled zone is the spectral range considered in case $3$, while the red filled zone marks the spectral range considered in cases $1$ and $2$ (upper right and lower right panels, respectively).
              }
         \label{fig:confroptuv}
   \end{figure*}
%------------------
In this section, we present a more generic test to explore the difference in the capability of retrieving the stellar metallicity of galaxies when using a small set of indices at high S/N ($\ge 20$) or a larger one at a moderate S/N ($= 10$). In particular, we compared the results obtained using the classic set of optical indices ([$H\beta$, Mgb, Fe$5270$, Fe$5335$, Fe$5406$]) with those obtained using a larger set of indices between $3500$ and $5000 \AA$. This test is not specifically related to StePS-like spectra, and aims to inform other possible future spectroscopic surveys. For this reason, we did not use WEAVE-StePS-like simulated spectra (i.e. with $S/N$ in the UV region depending on the spectral shape of the simulated galaxies for the same S/N fixed in the I-band region), but relied on a new set of ad hoc simulations, assigning a fixed S/N to the continuum adjacent to each spectral index considered.  We selected three different set of indices: (1) [H$\beta$, Mgb, Fe$5270$, Fe$5335$, Fe$5406$], that is, the so-called "classic" indices, (2) [H$\beta$, Mgb, Fe$5270$] and (3) [BL$3580$, Fe$3619$, H$\delta_\text{F}$, H$\gamma_\text{F}$, G$_\text{band}4300$, Fe$4383$, Ca$4455$, Fe$4531$, and H$\beta$], the first two at wavelengths larger than $4800\AA$, while the last one at wavelengths lower than $5000\AA$. We want to test whether it is more profitable for the stellar metallicity estimates to invest in observing time to obtain S/N = 20-30 around the first two set of indices or to obtain a S/N = 10 around the larger and bluer set of indices.
The upper panel of Figure~\ref{fig:confroptuv} shows the comparison of the results obtained using the five "classic" redder indices (i.e. case $1$) at S/N = 20 (red error bars) and the larger set of $9$ bluer indices (case $3$) at a lower S/N of $10$ (black error bars). It is clear that doubling the number of indices at lower S/N at $3500\AA <\lambda < 5000\AA$ gives similar results than using only the five classic indices at higher S/N ($4600\AA <\lambda < 5600\AA$). Moreover, the use of bluer indices allows us to obtain a better metallicity constraint at younger ages (as demonstrated by the black error bars in Fig.~\ref{fig:confroptuv}) becoming smaller than the red ones at ages younger than $1$ Gyr. %It is worthy to note that, from an observational point of view, $S/N = 10$ in the UV region corresponds roughly to $S/N = 20$ in the optical one for D$_n4000 \sim 1.3$ (see \cite{costantin2019few}).
The lower panel of Figure~\ref{fig:confroptuv} shows the comparison of the results obtained using even a smaller set of three "classic" redder indices (case $2$, $4600\AA <\lambda < 5400\AA$) at S/N = 30 (red error bars) and the full set of $9$ bluer indices (case $3$) at S/N = 10 (black error bars). As in the upper panel, we found that results similar to those obtained with few optical indices at a relative high S/N can be obtained with a wider set of bluer indices, over a larger wavelength baseline, for spectra at a lower S/N. In this second comparison, the improvement on the metallicity estimates is higher, in particular at young ages when using the larger and bluer set of indices at S/N = 10, namely, one-third of the S/N value considered for the redder indices only. This result is in agreement with what previously found by \cite{lopez2016simultaneous}. They obtained similar constraints of the stellar metallicity combining optical spectra with FUV and NUV photometric data at S/N = 10 than using only optical spectra at S/N =20.

Our results show that metallicity can still be reliably measured with optical and blue spectral indices even at relatively low S/N values.
Indeed, this is relevant when also taking into consideration that bluer indices are less affected by the sky emissions at $\lambda > 6000\AA$ up to $z \sim 1$, whereas above $z \sim 0.5,$ the redder indices are significantly affected by strong sky emission lines and telluric absorptions. This results could be useful for wavelength range selection in planning future spectroscopic surveys.

While our analysis shows that blue indices have a significant constraining power to metallicity, we caution that generally, they are also sensitive to the effect of non-solar abundance ratios \citep[see e.g.][]{vazdekis2015evolutionary}, and therefore it would be ideal  to estimate metallicity with both optical and UV features if modelling based on different chemical composition is not available.% In any case, the effect of non-solar abundances becomes less and less important at younger ages, i.e. at increasing redshift.

%------------------

\section{Summary and conclusions}
\label{sec:conclusion}
In this paper we investigate the capability to retrieve the mass-weighted metallicity in galaxies at different redshifts and S/N values, simultaneously exploiting the UV and optical rest-frame wavelength coverage. Our analysis is based on a wide stellar templates library derived from the latest version of the \cite{bruzual2003stellar} models. We showed that the mid-UV indices BL$3580$ and Fe$3619$ can provide reliable constraints on stellar metallicity, when used along with other optical indicators. At the same time, we emphasised that UV indices below $3000\AA$  can be strongly affected by the presence of even a tiny fraction (i.e. < $0.1\%$ of the overall mass) of very young (i.e. age $< 100$ Myr) stars. The same effect is expected from the presence of old hot stars (e.g. PAGB/BHB stars) which can mimic the same behaviour of very young ones \citep[e.g.][]{salvador2020sub,salvador2021young,salvador2022lessons}. Nevertheless UV spectral indices can be used to infer the metallicity of the stellar populations in galaxies, in particular considering the spectral region around $\sim 3500\AA$. 

To analyse the expected constraining power of different sets of indices on real observations, we simulated spectra as will be observed by the upcoming WEAVE-StePS survey. This intermediate redshift survey will observe around $25000$ galaxies at $0.3<z<0.7$ covering a wide spectral range. Therefore, we simulated $25000$ spectra at three different redshift ($z = [0.30,0.55,0.70]$) and at four values of S/N in the observed I-band $S/N_{I,\textrm{obs}} = [10,20,30,50]$.
We focused on the constraint of the metallicity by measuring key UV and optical absorption-line indices for each simulated galaxy and using a Bayesian approach with an extended galaxy templates library to retrieve the mass-weighted metallicity with an accurate estimate of the expected observational errors. We found that, in general, the available spectral indices can be used to reliably constrain stellar metallicity already at $S/N_{I,\textrm{obs}} = 10$, with $\sigma \le 0.3$ dex for galaxies older than $\sim 2$ Gyr. Below this age, metallicity indices are weak, and mostly sensitive to the temperature of the stars, namely, to the age of the stellar population. 

%We showed that reliable metallicity estimates can already be obtained at $S/N_{I,\textrm{obs}} = 10$, with better constraints as age increases, following the age-metallicity degeneracy. The median errors decrease with the age, with values from $\sigma = 0.38$ dex for $age_{bin} = 0.40$ Gyr to $\sigma = 0.26$ dex for $age_{bin} = 5$ Gyr, with no systematic deviations at any galaxy age.
Our results are in good agreement with other theoretical \citep{choi2014assembly} and observational \citep{gallazzi2005ages} results in literature, also demonstrating how metallicity estimates can take advantage of including bluer and UV indices together with the classic optical ones.

Finally, we performed a direct comparison among different sets of indices when characterising the stellar metallicity to show the importance and the efficiency of using bluer and UV indices. For the comparison to be fair, we relied on a new set of simulated spectra with fixed S/N along the continuum around each spectral index. We found that results similar to those obtained with fewer optical (e.g.  [H$\beta$, Mgb, Fe$5270$]) indices at relative high S/N can be obtained with a wider set of bluer indices ([BL$3580$, Fe$3619$, H$\delta_\text{F}$, H$\gamma_\text{F}$, G$_\text{band}4300$, Fe$4383$, Ca$4455$, Fe$4531$, H$\beta$]), over a larger wavelength baseline, at lower S/N. We emphasised that one main advantage of blue spectral indices is that of avoiding wavelength regions, in the galaxy observed frames, that are strongly affected by atmospheric contamination (both emission lines and telluric lines) at intermediate redshift. However, blue indices are expected to be more affected by the effect of non-solar abundance ratios, though the latter become less and less important at young ages (i.e. higher redshifts).

This work has demonstrated the good level of accuracy that can be reached when measuring the stellar metallicity in galaxies even at quite low S/N if a large number of indices can be employed, including (in particular) some UV indices above $3000\AA$. This is very a promising result for the upcoming surveys with new, high multiplexed, large field spectrographs, such as StePS at the WEAVE and 4MOST instruments, which will provide spectra of thousands of galaxies covering large spectral ranges (between $3600$ and $9000\AA$) at $S/N> 10\AA^{-1}$. 

%This work was established as a feasibility study for the WEAVE-StePS survey to investigate the capability to infer metallicity of the stellar populations, showing promising results. Naturally, our analysis can be adapted to studies based on data coming from other similar instruments and surveys, with adequate $S/N$ and a wide wavelength range, such as the future $4$MOST-StePS. In particular, the $S/N$ that this survey will reach will permit to obtain reliable measurements of metallicity in individual galaxies with an accuracy suitable for the study of the galaxy evolution.

\begin{acknowledgements}
      F.R.D., A.I., M.L, S.Z., A.G., F.L.B. acknowledge financial support from grant 1.05.01.86.16 - Mainstream 2020. A.F.M acknowledges support from RYC2021-031099-I and PID2021-123313NA-I00 of MICIN/AEI/10.13039/501100011033/FEDER,UE. L.C. acknowledges financial support from Comunidad de Madrid under Atraccion de Talento grant 2018-T2/TIC-11612 and Spanish Ministerio de Ciencia e Innovacion MCIN/AEI/10.13039/501100011033 through grant PGC2018-093499-B-I00. R.G.B. acknowledges financial support from the grants CEX2021-001131-S funded by MCIN/AEI/10.13039/501100011033 and to PID2019-109067-GB100. AV acknowledges support from grant PID2019-107427GB-C32 and PID2021-123313NA-I00 from the Spanish Ministry of Science, Innovation and Universities MCIU. This work has also been supported through the IAC project TRACES, which is partially supported through the state budget and the regional budget of the Consejer{\'{i}}a de Econom{\'{i}}a, Industria, Comercio y Conocimiento of the Canary Islands Autonomous Community. AV also acknowledges support from the ACIISI, Consejer{\'{i}}a de Econom{\'{i}}a, Conocimiento y Empleo del Gobierno de Canarias and the European Regional Development Fund (ERDF) under grant with reference ProID2021010079.
\end{acknowledgements}


\bibliographystyle{aa}
\bibliography{biblio}

\begin{appendix} %First appendix
\section{$\alpha$-enhancement}
\label{sec:appA}
%----------------------------------------------------------------- 
   \begin{figure*}
   \centering
   \includegraphics[width=0.95\textwidth]{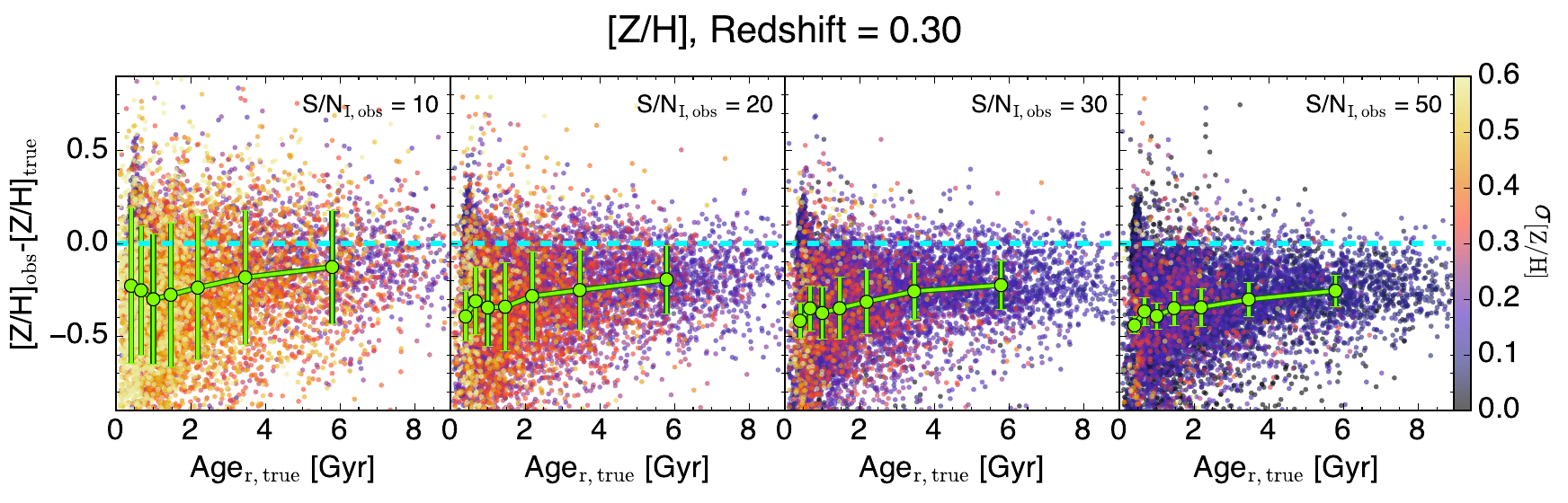}
      \caption{Difference between the observed and true metallicity, as a function of true $r$-band light-weighted age values at $z = 0.30$, considering $\alpha$-enhanced simulations. Each dot is colour-coded according to $\sigma_\textrm{[Z/H]}$ obtained from the posterior. The green error bars are the median and the $68\%$ confidence interval of [Z/H] differences in bins of Age$_{r,\text{true}}$, with each bin having the same number of data points. The cyan dashed line represents the $0$ value of the y-axis.
              }
         \label{fig:scatteralfaage}
   \end{figure*}
Our analysis relies on stellar population models constructed with stars having the same abundance pattern as in the solar neighbourhood. Since massive galaxies are over-abundant in several elements (e.g. [$\alpha$/Fe]), we estimate the possible impact of [$\alpha$/Fe] on metallicity estimates by adding the [$\alpha$/Fe] information on the simulated spectra, while the comparison library remains unaltered. We consider the most extreme case of [$\alpha$/Fe]=0.4, which is typical of the most massive galaxies in the nearby Universe. We added the [$\alpha$/Fe] information in a differential way, by measuring the indices used in this work on [$\alpha$/Fe] $= 0.4$ and [$\alpha$/Fe] $= 0$ templates and then we subtract the values of the [$\alpha$/Fe] $= 0$ indices to the ones at [$\alpha$/Fe] $= 0.4$,  to obtain the differential offset caused by the $\alpha$/Fe. The differential offset is added on the indices values of the WEAVE-StePS-like simulations. For the UV part (below $3500\AA$) we used the E-MILES preliminary models \citep[see][]{eftekhari2022strong} calculated with Teramo isochrones, age values between $4$ and $10$ Gyr, metallicity values between $-0.35$ dex and $0.26$ dex, with [$\alpha$/Fe] $= 0$ and [$\alpha$/Fe] $= 0.4$, respectively. For the optical part (above $3500\AA$) we used the $\alpha$-enhanced E-MILES models from \cite{vazdekis2015evolutionary}. For ages below $4$ Gyr and metallicity below $-0.35$ dex we consider the differential offset obtained at $4$ Gyr and at $-0.35$ dex, as no $\alpha$-enhanced models have been computed in the UV spectral range.

Figure~\ref{fig:scatteralfaage} shows the difference between the estimated metallicity and the true ones, as a function of the true values of $r$-band light-weighted age, at $z = 0.30$, considering $\alpha$-enhanced simulations. There is a systematic underestimation up to $-0.3$ dex at $S/N_{I,\textrm{obs}} = 10$, while the errors decrease as the $S/N_{I,\textrm{obs}}$ increases. The underestimation is large because our metallicity indices mostly consist of Fe indicators, which decrease as $\alpha$/Fe increases. While Figure~\ref{fig:scatteralfaage} shows the importance of taking the effect of abundance ratios into account, in practice, the effect of chemical abundances might be not so severe, as the overabundance of different elements (e.g. [C/Fe]) tend to cancel out the effect of [$\alpha$/Fe] \citep[see][]{la2017imf}.

%------------------

\end{appendix}

\end{document}